\documentclass[usenatbib]{mn2e}
\usepackage{graphicx,times}
\usepackage{bm}
\usepackage{epsf}

\usepackage{bm}
\usepackage{amsmath}
\usepackage{amssymb}

\usepackage{color}
\def\cred{\textcolor{black}}
\def\cg{\textcolor{black}}

\def\rtf{r,\theta,\varphi}
\def\br{{\bf r}}

\def\n{\noindent}
\def\qtwo{\qquad\qquad}

\def\myC{{\cred{\cal C}}}
\def\myD{{\cred{\cal D}}}

\def\ho{{\hat \Omega}}

\def\bbK{{\rm K}}

\def\2p{{(2\pi)^2}}

\def\be{\begin{equation}}
\def\ee{\end{equation}}

\def\ben{\begin{eqnarray}}
\def\een{\end{eqnarray}}

\def\beqa{\begin{eqnarray}}
\def\eeqa{\end{eqnarray}}

\def\oh{{\hat\Omega}}
\def\nn{{\nonumber}}

\def\inte{\int_0^{\infty}}

\newcommand{\edth}{\,\eth\,}

\def\suml{\sum_{l=0}^{l=\infty}}
\def\summ{\sum_{m=-l}^{m=l}}

\def\y8y{{}_sY^*_{lm}(\thet,\phi)}
\def\zz{{}_2Z_{klm}(\br)}

\def\szstar{{}_{s}Z^*_{klm}(\br)}
\def\szstarp{{}_{s'}Z^*_{k'l'm'}(\br)}

\def\sz{{}_sZ_{klm}(\br)}

\def\gg{{}_s\Gamma}
\def\ggp{{}_{s'}\Gamma'}
\def\ggpp{{}_{s''}\Gamma''}
\def\ggppp{{}_{s'''}\Gamma'''}

\def\sg{\Gamma}
\def\sgp{\Gamma'}
\def\sgpp{\Gamma''}
\def\sgppp{\Gamma'''}

\date{\today,~ $ $Revision: 0.9 $ $}

\begin{document}

\onecolumn

\title[Three-dimensional Statistics of Weak Lensing Shear and Flexion]
{Higher Order Statistics for Three-dimensional Shear and Flexion}

\author[D.Munshi et al.]
{Dipak Munshi$^{(1)}$, Thomas Kitching$^{(2)}$, Alan Heavens$^{(2)}$, Peter Coles$^{(1)}$ \\
$^{(1)}$ School of Physics and Astronomy, Cardiff University, Queen's
Buildings, 5 The Parade, Cardiff, CF24 3AA, UK\\
$^{2}$Scottish Universities Physics Alliance (SUPA), Institute for Astronomy,
University of Edinburgh, Blackford Hill,  Edinburgh EH9 3HJ, UK }

\maketitle

\begin{abstract}
We introduce a collection of statistics appropriate for the study of
spinorial quantities defined in three dimensions, focussing on
applications to cosmological weak gravitational lensing studies in
3D. In particular, we concentrate on power spectra associated with
three- and four-point statistics, which have the advantage of
compressing a large number of typically very noisy modes into a
convenient data set.  It has been shown previously by \cite{MuHe09}
that, for non--Gaussianity studies in the microwave background, such
compression can be lossless for certain purposes, so we expect the
statistics we define here to capture the bulk of the cosmological
information available in these higher-order statistics. We consider
the effects of a sky mask and noise, and use Limber's approximation
to show how, for high-frequency angular modes, confrontation of the
statistics with theory can be achieved efficiently and accurately.
We focus on scalar and spinorial fields including convergence, shear
and flexion of 3D weak lensing, but many of the results apply for
general spin fields.

\end{abstract}

\section{Introduction}

In this paper, we consider a set of new statistics designed to
encapsulate much of the information content of third-order and
higher statistics for spinorial fields defined in three dimensions.
In cosmological applications such higher-order statistics can be
very noisy, and the dimensionality of the space may also lead to a
very large number of data to consider.  Thus some form of data
compression is attractive, but preferably in a way which does not
reduce the cosmological information content inherent in the original
statistics.  In this paper, we build on the ideas originally
presented in \cite{MuHe09}, where it was shown that one statistic,
the power spectrum associated with the bispectrum, could be used
very effectively to estimate non-gaussianity in the microwave
background radiation, as it was a lossless compression for this
purpose, and also had the important added benefit of being able to
provide evidence that a non-gaussianity is primordial.  In this
paper, we extend the ideas to cover spin-weighted fields which are
defined in three dimensions, with particular emphasis on weak
lensing fields convergence, shear and flexion.

Weak gravitational lensing of background source galaxies is caused by
fluctuations in the intervening mass distribution. It manifests itself in a number of ways, most notably as
distortions in their images. This effect arises due to the fluctuations
of the gravitational potential and consequent deflection of light by
gravity.

Despite being a relatively young subject weak  gravitational lensing
\citep{MuPhysRep08}  has made major progress within the last decade,
since the first measurements were published
\citep{BRE00,Wittman00,KWL00,Waerbeke00}. There has been
considerable progress in analytical modelling, technical
specification and the control of systematics. By its dependence on
the mass power spectrum at lower redshifts, weak lensing surveys
play a complementary role to the studies based on large-scale galaxy
surveys and Cosmic Microwave Background (CMB) observations. Ongoing
and future weak lensing surveys such as the CFHT legacy
survey{\footnote{http://www.cfht.hawaii.edu/Sciences/CFHTLS/}},
Pan-STARRS  {\footnote{http://pan-starrs.ifa.hawaii.edu/}}, the Dark
Energy Survey,  and further in the future, the Large Synoptic Survey
Telescope {\footnote{http://www.lsst.org/lsst\_home.shtml}}, WFIRST
{\footnote{http://wfirst.gsfc.nasa.gov/}} and Euclid
{\footnote{http://sci.esa.int/euclid}} will provide a wealth of
information in terms of mapping the distribution of mass and energy
in the universe.

Owing to the lack of photometric redshift information the
traditional approach to weak lensing has largely adopted a 2D
approach, analysing correlations of the shapes of galaxy images on
the sky only. However the availability of photometric redshifts
allows a 3D weak lensing analysis, which was introduced  by
\citet{Heav03}. Later developments by various authors \citep{HRH00,
HKT06, HKV07, Castro05} have shown that it can play a vital role in
constraining the dark energy equation of state \citep{HKT06} and the
neutrino mass \citep{Kit08}. This has lead to recent progress in
modelling weak lensing observables in 3D extending results
previously obtained in projection or using tomographic techniques
\citep{MuHeCo_wl1_10}.


Early results on analytical modelling typically assumed a small
survey size and adopted a 2D approach that uses a flat-sky
formalism. This is related to the fact that first generation of
surveys typically covered a small portion of the sky and lacked any
redshift information \citep{JSW00}. Indeed such analytical modelling
was very successful in predicting lower-order statistical properties
of weak lensing convergence and shear very accurately
\citep{MuJai01,Mu00,MuJa00}. These results depends on analytical
modelling of underlying density perturbations using perturbative and
empirical methods. \citep{MuJai01,Valageas00,
MuVa05,VaMuBa04,VaMuBa05}. A tomographic step was next advocated to
tighten the cosmological constraints. The tomographic studies
typically divides the sources into a few redshift slices
\citep{Hu99,TakadaWhite03,TakadaJain04,Massey07,Schrabback09}. These
slices are then analyzed essentially using a two-dimensional
approach but including the correlation between different redshift
slices. A notable exception to the 2D analysis was
\citet{Stebbins96} who developed an all-sky formalism for weak
lensing surveys. The techniques developed in \citet{Stebbins96} rely
on a tensorial formalism,  whereas we will be using an equivalent
treatment based on spin weight spherical harmonics. Extending
previous studies by \citet{Heav03} and \citet{Castro05},
\cite{Mu_wl2_10} extended the all-sky formalism to 3D to take into
account the photometric redshift information, as well as extending
to higher-order statistics. However they focused on the convergence
field which, being a spin-$0$ field, is relatively easier to
analyze. The main motivation behind this work is to extend previous
results to arbitrary spinorial fields such as shear and their
derivatives flexion.


Weak lensing at small angular scales probes the nonlinear regime of gravitational
clustering, and the extra modes there can lift degeneracies about
background cosmology present in studies involving the power spectrum
alone see, e.g., \cite{BerVanMell97,JainSeljak97, Hui99,Schneider98,TakadaJain03}.
The nonlinear regime is characterized  by gravity-induced non-Gaussianity, and detailed studies that employ the Fisher matrix formalism have already
demonstrated the potential of using higher-order non-Gaussianity
information to lift cosmological degeneracies.
Higher-order studies are also important in evaluating the variance of
lower-order statistics, e.g. a proper knowledge of the trispectrum
is essential for computing the error bars in the power spectrum\citep{TakadaJain09}. The modelling of higher-order
statistics typically involves either perturbative
techniques or empirical modelling of the underlying matter clustering
\citep{Fry84,Schaeffer84, BerSch92,SzaSza93, SzaSza97, MuBaMeSch99,
MuCoMe99a, MuCoMe99b, MuMeCo99, CMM99, MuCo00, MuCo02, MuCo03}.
Using such prescriptions and their extensions, studies involving non-Gaussianity,
have also been performed in projection (2D) as well as using
tomographic information \citep{Hu99,TakadaJain04,TakadaJain03,Semboloni08}
with remarkable success.

Studies involving higher-order correlation functions have been
performed using observational data \citep{BerVanMell97,BerVanMell02,Pen03,JBJ}.
Most of these studies involve one-point moments (cumulants) which collapse
the entire correlation function into a single number.
Mode-by-mode estimates of higher-order correlation functions or
multispectra though far more interesting is difficult given the
low signal-to-noise of current observational data. Current studies
by \cite{MuHe09} defined power spectra associated with each multispectrum
that uses an intermediate option in data compression.
While initially this concept was applied to CMB studies, recent work by
\citep{MuHeCo_wl1_10} extended this concept to weak lensing. This
initial work focused on convergence $\kappa$. Being a spin-$0$ (scalar) object,
the analysis of convergence statistics is relatively simple. In their analysis \cite{Mu_wl2_10}
used the similar statistics for shear and flexion fields but in projection (2D). Th main motivation
for the present study is to use the full 3D information (available from photometric redshift surveys) in analyzing
the non-Gaussianity not only in the convergence field but also in shear and flexion. This is particularly interesting
as current photometric redshift surveys with good image quality will provide a wealth
of data for the analysis of weak lensing which can be used to probe cosmological information.
For our study, we combine well-motivated ansatz\'e in modelling the gravitational clustering
with the Limber approximation. The results that we derive here are
generic and will be useful in other areas of cosmology where
integration along line of sight is involved. To keep the results simpler
we will ignore the fact, that in a realistic survey, the average density of sources
will decline with distance, and the distance estimated from photometry will also
include error, but these are evidently important ingredients in a practical implementation of these statistics.


The expressions for higher-order multispectra generically include
multidimensional integrals involving multiple spherical Bessel functions.
We will be using Limber approximation to simplify these results.
We will show that, at each order, we can reduce the dimensionality of these integrals to unity by using
Limber approximation. This will simplify the numerical evaluations
of such integrals considerably.

This paper is arranged as follows. In \textsection2 we discuss the basic formalism
of 3D weak lensing. The formalism presented here is a generalization of
\citep{MuHeCo_wl1_10} and \citep{Mu_wl2_10} and can analyze higher-order statistics
of spinorial fields in 3D.
In \citep{MuHeCo_wl1_10} results were derived for higher-order statistics for the convergence
and in \citep{Mu_wl2_10} the focus was on higher-order statistics of spinorial objects but in projection (2D).
The notations for 3D harmonic decomposition, which will be used in the following sections are also introduced
here. In \textsection3 we introduce the
models describing higher-order clustering of underlying matter which are then used to construct models for
the bispectrum and trispectrum in the nonlinear regime. The results obtained
are generic and can describe higher-order statistics of weak lensing convergence, shear and flexions.
In \textsection4 we focus on power spectra associated with higher-order multispectra.
Results presented in this section correspond to both all-sky and patch-sky coverage. In \textsection5
we focus on error analysis and derive results for scatter (or variance) of various estimators in
the presence of observational noise and mask. Finally \textsection6 is devoted to discussion of the results.
Though we have mainly focused on weak-lensing, the general formalism developed in the paper will have wider applicability.
We will use the Hierarchical ansatz to model clustering of underlying mass distribution, but the
treatment can also be adopted in the context of more elaborate scenarios of clustering e.g. halo model.

\section{Notation}

This section is devoted to introducing the basic notation and
formalism of 3D weak lensing.  We will  follow the notation used in
\citet{MuHeCo_wl1_10} which is based mainly on \citet{Heav03} and
further developed by \citet{Castro05}. The results of
\citet{Castro05} were generalized by \citet{MuHeCo_wl1_10} to take
into account higher-order correlations. The aim of this paper is to
extend both \citet{MuHeCo_wl1_10} and \citet{Mu_wl2_10} to the
analysis of shear using full 3D information.

\subsection{A Tale of Two Potentials}

Linking the 3D lensing potential $\phi$  and the 3D gravitational
potential $\Phi$ is crucial in connecting lensing observables to
theory.  In this subsection we will consider the harmonic
decomposition of 3D scalar (spin-0) fields which is a step towards
making this connection because examples of such fields include the
scalar potentials and the convergence field $\kappa$ that we
encounter in weak lensing. The harmonic decomposition is most
naturally done using eigenfunctions that can be constructed using
ordinary spherical harmonics and spherical Bessel functions. In the
next subsection we will generalize them to the case of spinorial
fields.

The statistics of shear and convergence can be expressed in a
natural way through their relation to $\Phi(r,\theta,\varphi)$ the
3D gravitational potential at a 3D position $\rtf$, and  $\phi(\br)$
the lensing potential. The density contrast $\delta(\br)$ is
directly related to the potential through the Poisson equation. This
allows us to link directly  the statistics of the weak lensing
observables to the underlying statistics of the mass distribution,
and hence to cosmological parameters. The radial distance $r(t)$ is
related to the Hubble expansion parameter $H(t)= \dot a/a$ by $r(z)
= c\int_0^{z}dz' / H(z')$. The Hubble parameter is sensitive to the
contents of the Universe thereby making weak lensing a useful probe
to study dark energy. The line of sight integral relating the two
potentials can be written as \citep{Kaiser92}: \be \phi(\br) \equiv
\phi(r,\ho) = {2 \over c^2} \int_0^r dr' F_{\rm K}(r,r')
\Phi(r',\ho); \qquad F_{\rm K}(r,r')\equiv { f_{\rm K}(r-r') \over [
f_{\rm K}(r) f_{\rm K}(r')]}. \ee The Born approximation was used to
derive the above expression
\citep{BerVanMell97,Schneider98,Waerbeke02} The lensing potential
$\phi(\br) \equiv \phi(r,\ho)$ has a radial dependence and is a 3D
quantity. In our notation $r=r(t)$  is the comoving distance to the
source whose observed light was emitted at a given instance of time
$t$. The observer is situated at the origin. The function  $F_{\rm
K}(r,r')$ depends on the background cosmology. through the function
$f_{\rm K}(r)$; $f_{\rm K}(r) = \sin r, r, \sinh r$  for a closed
$({\rm K}=1)$, flat $({\rm K}=0)$ or open $({\rm K}=-1)$ universes
respectively.  Our convention for the Fourier transform for the 3D
fields is as in \cite{MuHeCo_wl1_10}. The eigenfunctions of the
Laplacian operator in flat space when expressed in spherical
coordinates turn out to be a product of spherical Bessel functions
$j_l(kr)$ in the radial direction and the spherical harmonics on the
surface of a unit sphere i.e. $Y_{lm}(\ho)=Y_{lm}(\theta,\phi)$. The
eigenfunctions $Z_{klm}(\br)=\sqrt{2 \over \pi}\,k
\,j_l(kr)\,Y_{lm}(\oh)$ are associated with eigenvalues $-k^2$. In
general the radial eigenfunctions are ultra-spherical Bessel
functions, but they can be approximated by spherical Bessel
functions when the curvature is small. The eigendecomposition and
its inverse transformation can be expressed as: \be \Phi_{lm}(k) =
\int d^3\br \Phi({\bf r}) Z_{klm}(\br); \ee and \be \Phi({\bf r}) =
\sum_{l=0}^{\infty} \sum_{m=-l}^{m=l} \, \int dk \,
\Phi_{lm}(k)Z_{klm}(\br). \ee

The specific choice of eigenfunctions allows allows us easily to express the coefficients of expansion of the
convergence (or shear) in terms of the expansion of the density field through the Poisson equation \citep{Heav03}
$\triangle \Phi(\br) = {3} \Omega_m H_0^2 \delta(\br)\,/2a$. In the harmonic domain this can be
expressed as $\Phi_{lm}(k;r) ={A}\, \delta_{lm}(k;r)/a(r)\,k^2$ with  $\cred{A \equiv -{3 \Omega_m H_0^2/2}}$.
Here, $\Phi_{lm}(k)$ is the spherical harmonic decomposition
of $\Phi({\bf r})$, and similarly for $\phi({\bf r})$.
In our notations,  $a(z)=1/(1+z)$ is the scale factor at redshift $z$,
$\Omega_m$ is the total matter density at $z=0$, and $H_0$ is the
Hubble constant today. $\delta_{lm}(k;r)$ is the eigendecomposition
of $\delta(\br)$. When appearing after the semi-colon, the $r$ dependence (e.g. of $\Phi_{lm}(k;r)$) is really an
expression of the time-dependence of the potentials, which translates to a dependence on $r$, as $r$ depends on look-back time.
Using these decompositions, the harmonic decomposition of the lensing
potential $\phi_{lm}(k)$ and the 3D gravitational potential $\Phi_{lm}(k,r)$ are related by the
following expression \citep{Castro05}:

\be \phi_{lm}(k) = {4 k \over \pi c^2 } \int_0^\infty dk' k'
\int_0^{\infty} r\cg{^2} dr j_l(kr) \int_0^r dr' \cg{F_K(r,r')}
j_l(k'r')\Phi_{lm}(k';r'). \label{Den} \ee The basis functions for
the harmonic decomposition of the spinorial fields such as flexion
and shear will involve spin-weight spherical harmonics which we will
introduce next. The 3D power spectra for the gravitational
potentials $\Phi$ and the lensing potential $\phi$ are defined
through the following expressions: \be \langle
\Phi_{lm}(k)\Phi_{l'm'}(k') \rangle =
\myC_l^{\Phi\Phi}(k)\delta_{1D}(k+k')\delta^K_{ll'}\delta^K_{mm'};
\quad \langle \phi_{lm}(k)\phi_{l'm'}(k') \rangle =
\myC_l^{\phi\phi}(k,k')\delta^K_{ll'}\delta^K_{mm'}.\ee

\subsection{3D Eigendecomposition of Spinorial Functions}

In this subsection we will introduce the generic spin-weight functions and their eigendecomposition.
Specific cases that are of interest here include shear and flexions.
This will generalize the spin-0 results discussed above for the convergence field. We can
expand the fields such  as shear $\gamma_{\pm}(\br)$, flexions ${\cal F}(\br), {\cal G}(\br)$
in 3D basis functions that are constructed out of spin-weight spherical harmonics $_sY_{lm}(\oh)$
on the celestial sphere and spherical Bessel functions $j_l(kr)$  in the radial direction.
Expansion in such bases provides a very simple relationship between harmonic coefficients
of the shear, flexion and convergence on the one hand, and the lensing potentials on the other.
Moreover, spherical coordinates are the natural choice for eigendecomposition as this
provides a clear separation in terms of radial modes and the modes on the
surface of the sky, and the ubiquitous presence of a sky mask induces mixing of modes only on the
surface of the sphere, and the use of photometric redshift estimates
only introduces error in the radial direction without altering the angular position.
The choice of eigenfunction is also motivated by the Poisson equation
which relates the 3D potential $\Phi(\br)$ to the density distribution
$\delta(\br)$, whose statistical property we will model to predict
the statistics of shear, convergence or flexion. Extending the definition of
spin-$0$ eigenfunctions $Z_{klm}(\br)$ we will denote the spin-$s$ eigenfunctions as $_sZ_{klm}(\br)$; which is defined as: ${}_sZ_{klm}(\br) = \sqrt {2 \over \pi} k j_l(kr) {}_sY_{lm}(\oh)$.
The spin-weight spherical harmonics are defined in terms of D-matrices \citep{VMK88,PenRind84}.
They satisfy a orthogonality relationship similar to ordinary spherical
harmonics. Spin-weight harmonics with the same or different spin indices are orthogonal
on the surface of sky. This generalizes the orthogonality relationship
of ordinary spin-zero spherical harmonics: ${}_sY_{lm}(\oh) = \sqrt{2l+1 \over 4\pi } D^l_{-s,m}(\theta,\phi,0)$,
$\int~~d\oh~_{s}Y_{lm}(\oh) _{s'}Y^*_{l'm'}(\oh)d\oh
= \delta_{ll'}\delta_{mm'}\delta_{ss'}$.

Alternative expansion schemes are indeed possible such as using
tensor spherical harmonics but they are perhaps more difficult to
work with. It is worth noting here that the formalism of
spin-harmonics is extensively used in studies involving Cosmic
Microwave Background Polarization \citep{Bunn03}. The forward and
inverse transform of an arbitrary spin function ${}_sf(\br)$ from
real space to harmonic space links it with its harmonic components
$_sf_{lm}(k)$ that can be expressed as: ${}_sf(\br) =
\int_0^{\infty} dk \suml \summ [_sf_{lm}(k)]\sz$ and $_sf_{lm}(k) =
\int d^3\br [{}_sf(\br)] \szstar$. The orthogonality relationship
satisfied by the 3D spherical basis functions $[\szstar]$ depends on
the orthogonality of spin-weight harmonics $_sY_{lm}(\oh)$ and that
of the spherical Bessel functions $j_l(kr)$. It generalizes a
similar relation for the scalar harmonics. For arbitrary spinorial
fileds with spins $s$ and $s'$ it reads: $\int d^3\br
[\szstar][\szstarp] =
\delta_D(k-k')\delta_{ll'}\delta_{mm'}\delta_{ss'}$. The inverse
transforms are used to define the harmonic components of generic
spinorial fields $\eta(\br)$ and $\eta^*(\br)$. The results that we
will derive in our later sections are expressed most naturally in
the harmonic domain using these components ${}_2\eta_{lm}(k)$ and
${}_{-2}\eta_{lm}(k)$ which can be expressed as: ${}_2\eta_{lm}(k) =
\int d^3\br ~ \eta(\br) \zz$ and ${}_{-2}\eta_{lm}(k) = \int d^3 \br
~ \eta^*(\br) \zz$. It is indeed possible to work with
${}_2\eta_{lm}(k)$ and ${}_{-2}\eta_{lm}(k)$ as well as the
harmonics $E_{lm}(k)$ and $B_{lm}(k)$ that can be constructed from
them. Though they contain the same information, the {\em Electric}
or $E$ and {\em Magnetic} $B$ modes provides a rotationally
invariant description in full sky. The expansion coefficients
$E_{lm}$ has a parity $(-1)^l$ while $B_{lm}$ has a parity
$(-1)^{l+1}$. The clear separation of modes with different parity
gives a clear mathematical advantage in the case of weak lensing, as
It can be shown that, at first order, weak lensing from
gravitational clustering can only generate $E$ modes, whereas
systematics are mostly responsible for the generation of any $B$
mode contribution.

The explicit expressions for the electric $E_{lm}(k)$ and magnetic
$B_{lm}(k)$ components, constructed from these harmonic transforms
are: $E_{lm}(k) = -{1 \over
2}[{}_2\eta_{lm}(k)+{}_{-2}\eta_{lm}(k)]$ ; $B_{lm}(k) = {i \over
2}[{}_2\eta_{lm}(k)-{}_{-2}\eta_{lm}(k)]$; and ${}_{\pm 2}
\eta_{lm}(k) = -[E_{lm}(k) \pm i B_{lm}(k)]$. The individual
components of the field $\eta(\br)$, $\eta_1(\br)$ and $\eta_2(\br)$
are expressed in terms of eigenfunctions $Z_{+,klm}(\br)$ that can
be constructed from linear combinations of $Z_{\pm2,klm}(\br)$
introduced before. The formalism used here is very similar to
\cite{Mu_wl2_10}. The emphasis here however is not just on 2D
decomposition on the surface of the celestial sphere but rather on a
3D decomposition which relies on the photometric redshift to
estimate radial distance.

It is worth mentioning here that unique decomposition of a function
into modes $E$ and $B$ mode on the celestial sphere is possible only
with complete sky coverage. In the presence of a boundary, which is
often the case owing to the presence of masks, the decomposition is
ambiguous. For the case of weak lensing shear these equations can be
specialized further by ignoring the magnetic contribution which is
zero for shear generated purely by gravitational lensing in the
absence of any systematics. Indeed, higher-order lensing corrections
can generate lensing $B$ mode too \citep{CooHu02} but are
sub-dominant.

\begin{figure}
\begin{center}
{\epsfxsize=12.cm \epsfysize=6.cm {\epsfbox[30 430 588 712]{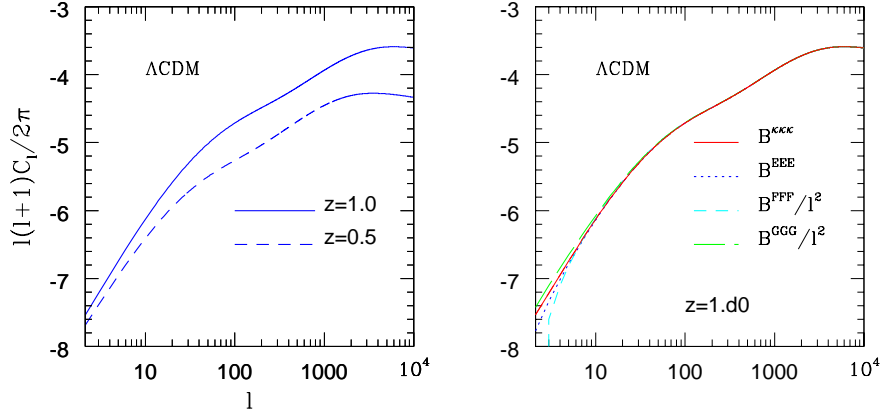}}}
\caption{Left panel compares the power spectrum $\myC_l$ for the convergence $\kappa$ for two
different source redshift $z_s=1$ and $z_s=0.5$. Right panel shows the power spectrum $\myC_l$ for
the convergence $\kappa$, electric part of the shear $E$, flexions $\cal F$ and $\cal G$
as a function of $l$. The flexion power spectra $\myC_l^{\cal F}, \myC_l^{\cal G}$ are
normalized by $l^2$ for display. The cosmology assumed is $\Lambda$CDM and all sources are
assumed to be at the same redshift of $z_s=1$.The $\Lambda$ CDM background cosmology that we have considered is
described by the following set of parameters: $\Omega_m=0.3$, $\Omega_\Lambda=0.7$, $\Gamma=0.21$, $h=0.7$ and $\sigma_8=0.90$.}
\label{fig:cl_ps}
\end{center}
\end{figure}

\subsection{Harmonic decomposition of Convergence, Shear and Flexion}

The results derived in previous section can directly be applied to
the case of flexions, shear and convergence. Most of the generic results
are applicable to the analysis of shear $\gamma$ if we specialize
the field $\eta$ with a spin$-2$ object and identify with 3D shear $\gamma$.
Complex shear $\gamma$ constructed from its individual components
$\gamma_{\pm}(\br) = \gamma_1(\br) \pm i \gamma_2(\br)$ acts as a
spin-$2$ object and can be expressed in terms of the
lensing potential $\phi$ using spin-derivatives (see \cite{Mu_wl2_10} and \cite{Castro05} for more
discussion on spin-derivatives) which are used to construct spinorial fields with
different spin-weights. The lensing potential plays the role
of the generic scalar field introduced earlier to express arbitrary spin functions.
We will use the same symbol $\phi$ for both. We will use the
generalized symbol ${}_s\Gamma$ for general spin fields which will
include products of shear fields as well as higher derivative spin
objects such as flexions. In our current notation ${}_2\Gamma =
\gamma$ and ${}_{-2}\Gamma = \gamma^*$: ${}_{2}\Gamma(\br) \equiv \gamma(\br) = {1 \over 2}\eth\eth [ \phi]$ and
$ {}_{-2}\Gamma(\br) \equiv \gamma^*({\bf r}) = {1 \over 2}{\bar\eth}{\bar\eth}[ \phi^*]$.
In general the scalar potential $\phi(\br)$ will have both electric $\phi_E(\br)$  and magnetic $\phi_B(\br)$ components:
$\phi(\br) =\phi_E(\br) + i\phi_B(\br)$.

The individual shear components $\gamma_1(\br)$ and $\gamma_2(\br)$ and the convergence $\kappa(\br)$
can be expressed in terms of a
complex lensing potential $\phi(\br) =\phi_E(\br) + i\phi_B(\br)$. As pointed out before
the magnetic part of the potential $\phi_B(\br)$ will take contribution mainly from systematics
and the electric part corresponds largely to pure lensing contribution $\gamma_1(\br)=
{1 \over 4}(\edth\edth+\bar\edth\bar\edth)\phi(\br)$; $\gamma_2(\br) = -{i\over 4}(\edth\edth - \bar\edth\bar\edth)\phi(\br)$
and $\kappa(\br) = {1 \over 4}(\edth\bar\edth+\edth\bar\edth)\phi(\br)$.

Derivatives of the shears are higher-spin objects.
Using these derivatives quantities such as {\em flexions}
are constructed, which are also often used in the context
of weak lensing studies (\citet{GN02,GB05,Bacon06, BG05,Schneider08}).
The two flexions that are most commonly used are also known as the first flexion  $\cal F$
(spin $-1$) and $\cal G$ which is also known as the second flexion (spin $3$).
These two flexions in combination can specify distortion
beyond what is described by shear. The flexions can be used
to describe weak ``arciness'' in images of lensed galaxies and their
relationship with the shapelet formalism is well documented
\citep{Ref03,BJ02,RefBac03}. The flexions $\cal F(\br)$ and $\cal G(\br)$ are both used
in the literature mainly for individual
halo profiles and also for the study of substructures
\citep{Bacon06}. We are mainly interested however in higher-order statistics of these objects for generic
underlying cosmological clustering. This is done by linking the
3D harmonic decompositions of the flexions to that of the
lensing potential $\phi(\br)$:  ${\cal F}(\br) = {1 \over 6} \left (\bar\edth\edth\edth+\edth\bar\edth\edth + \edth\edth\bar\edth \right)\phi(\br)$ and ${\cal G}(\br) = {1 \over 2}\bar\edth\bar\edth\bar\edth \phi(\br)$.
Flexions have been used primarily to measure the galaxy-galaxy lensing to probe the
galaxy halo density profiles. Their cosmological use will depend on an accurate
understanding  of gravitational clustering at small angular scales.

In Fourier space the harmonics of $\gamma(\br)$ and $\gamma^*(\br)$
can be expressed in terms of the harmonic coefficients of
$\Phi_E(\br)$ and $\Phi_B(\br)$ denoted by $E_{lm}(k)$ and
$B_{lm}(k)$ respectively: ${}_{\pm2} \Gamma_{lm}(k) = -[E_{lm}(k)
\pm i B_{lm}(k)]$. Analogously, the harmonics of $\cal F$ and $\cal
G$  denoted by ${\cal F}_{lm}(k)$ and ${\cal G}_{lm}(k)$ can also be
expressed in terms of the $\phi_{lm}(k)$. In the absence of B-modes
the harmonics of the shear components are directly related to the
harmonic component of the Electric field $E_{lm}$. The harmonic
transforms of the shear components and convergence  can also be
expressed in terms of the lensing potential $\phi$ as follows: $
\kappa_{lm}(k) = - {l(l+1)\over 2} \phi_{lm}(k)$; $\quad E_{lm}(k) =
-{1 \over 2} \sqrt{(l+2)!\over (l-2)!} \phi_{lm}(k)$; \;\; ${\cal
F}_{lm}(k) = {1 \over 6} l^{1/2}(l+1)^{1/2}\left ( 3l^2 + 3l-2
\right )\phi_{lm}(k)$; \;\; ${\cal G}_{lm}(k) = {1 \over 2} \sqrt
{(l+3)! \over (l-3)!} \phi_{lm}(k)$.

These harmonic expressions can be used to reconstitute the real
space spinorial fields: ${}_{\pm2} \Gamma(\br) =  \int k dk
\sum_{l=0}^{\infty} \sum_{m=-l}^{l} \sqrt{(l+2)! \over (l-2)! }
~\phi_{lm}(k) ~{}_{\pm2}Z_{klm}(r,\oh);$  \;\; ${\cal F}(\br) = \int
k dk \sum_{l=0}^{\infty} \sum_{m=-l}^{l} {\cal F}_{lm}(k)
~~_{-1}Z_{klm}(\br)$ \;\; and ${\cal G}(\br) = \int k dk
\sum_{l=0}^{\infty} \sum_{m=-l}^{l}  {\cal G}_{lm}(k)
~_{3}Z_{klm}(\br)$. These results derived above are useful in
linking the statistics of weak lensing fields $\kappa(\br)$
$\gamma(\br)$, $\cal F(\br)$ and $\cal G(\br)$ with those of the
underlying density field $\delta(\br)$ responsible for generation of
the lensing potential $\phi(\br)$ with the help of  Eq.(\ref{Den}).

The theoretical modelling of the underlying mass distribution
that we employ in our study is based on the hierarchical ansatz.
The hierarchical ansatz is more suited to model gravitational
clustering a smaller scales, which makes it particularly suitable
for modelling the flexion statistics which put more weight
on smaller scales. A comment about noise contribution due to intrinsic flexions
of source galaxies is in order. While it is relatively
easy to model the intrinsic ellipticity of source galaxies,
detailed modelling of intrinsic flexion of source galaxies
is much more complicated and depends heavily
on modelling of galaxy shapes beyond the simplest description.
This uncertainty is also expected to increase with survey depth.

In addition to shear, convergence and flexion which
are used in weak lensing studies, we can also consider a generic scalar
tracer field $\Psi$ in our study. Such fields can represent a suitable
large scale tracers which are sometimes used for cross-correlation
studies or studies involving weak lensing magnification.

The statistics of shear and flexions can be best related to that of convergence
with certain $l$ dependent multiplicative factors that we will call {\it form factors}.
In later sections $F_l^\sg$ will denote the form factor associated with
a generic spin-weight field $\sg$. So the form factor for the shear $\gamma_{\pm}$
will be denoted by $F^{\gamma_{\pm}}_l$.

\section{3D Weak Lensing Statistics: Power Spectrum and Beyond}

For the study of non--Gaussianity we need to go beyond the study of
power spectra. In this section we will present results for 2-, 3-
and 4-point statistics, and show the relation between observables
and theory.  The various multispectra involve multidimensional
integrals, which we simplify by employing various levels of
approximations involving the high $l$ behaviour of $j_l(x)$.

\subsection{Power spectrum}

\begin{figure}
\begin{center}
{\epsfxsize=6.cm \epsfysize=6.cm {\epsfbox[310 435 582 712]{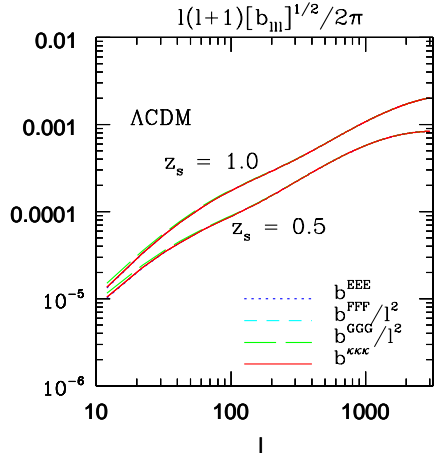}}}
\caption{The diagonal components of the reduced convergence bispectrum $b_{lll}$ are plotted as function
of $l$. The bispectrum is assumed to have a hierarchical form.
The hierarchical amplitude $Q_3$ is set to unity.
The source redshift is set to unity $z_s=1$. The plots for the flexions $\cal F$ and $\cal G$ are
normalized by $l^2$ for display. }
\label{fig:dia_red_bispec}
\end{center}
\end{figure}

We will start by deriving the power spectrum $C_l^{\sg\sgp}(k_1,k_2)$ for the 3D weak lensing fields.
Our derivation of the 3D convergence power spectrum is based on expressing its harmonic coefficients $\kappa_{lm}(k)$
in terms of the 3D density field  $\delta$ with the help of Poisson's equation and using the definition
of the convergence field in terms of the projected lensing potential $\phi$ gives:

\be
\kappa_{lm}(k) = {\cred{2} k A \over \pi c^2 }\cred{l(l+1)} \int_0^\infty dk' k' \int_0^{\infty} r^2 dr j_l(kr)
\int_0^r dr' F_\bbK(r,r')\, j_l(k'r'){\delta_{lm}(k';r') \over k'^2 a(r')}
\label{kappa2den}
\ee

We will use a shorthand notation $I_l(k_i,k)$ (defined below) useful
for simplification of our results. We will approximate the
cross-spectra at two different epoch using the approximation
$P^{\Phi\Phi}(k,r,r') =\sqrt{P^{\Phi\Phi}(k,r)P^{\Phi\Phi}(k,r')}$
\citep{Castro05}. Use of this approximation leads to separation of
respective integrals. As we will see below the use of the extended
Limber approximation, which is valid at high $l$, implies that
dominant contribution will come from single time slices $r=r'$ and
this approximation is not detrimental to any of the final results
which are quite generic. The 3D power spectrum can be expressed in
terms of $I_l(k_i,k)$ as \citet{Castro05}:

\ben
&& I_l(k_i,k) \equiv k_i \int_0^{\infty} dr~r^2~j_l(k_ir) \int_0^r dr'  F_\bbK(r,r')\, j_l(kr')
\sqrt{P^{\Phi\Phi}(k;r')} \nn \\
&& \myC_l^{\phi\phi}(k_1,k_2)= {16 \over \pi^2 c^{\cred{4}}}\int_0^{\infty} k^2 I_l(k_1, k)I_l(k_2,k)dk; \qquad
\myC_l^{\kappa\kappa}(k_1,k_2)= {1 \over 4} l^2(l+1)^2 \myC_l^{\phi\phi}(k_1,k_2); \qquad
C_l^{\sg\sgp}(k_1,k_2)= F_l^{\sg} F_l^{\sgp} C_l^{\kappa\kappa}(k_1,k_2).
\label{Cl_phi}
\een

\n Clearly the above expression is quite generic and contains all
the weak lensing information at the second-order level. This
expression is however quite cumbersome for any numerical
implementation as it involves three-dimensional integral which are
quite demanding computationally. We will be using extended Limber
approximation valid at high $l$ to simplify the above expression.
Using this approximation we can reduce the integrals to
one-dimensional integrals. In any case we will quote the generic
result that is valid without any approximation. Notice that the
following approximation is also independent of the factorization of
the power spectrum introduced before.

\ben
&& \myC_l^{\phi\phi}(k_1,k_2)  = {16 \over \pi^2 c^{\cred{4}}}\int k^2 dk I_l(k_1,k)I_l(k_2,k) \nn \\
&&  =  {16 \over \pi^2 c^{\cred{4}}} k_1 k_2 \int_0^\infty dr_a r_a^2 j_l(k_1r_a) \int_0^r dr_a' F_\bbK(r_a,r_a')
   \int_0^\infty dr_b r_b^2 j_l(k_1r_b) \int_0^r dr_b' F_\bbK(r_b,r_b')
\int k^2 dk ~P^{\Phi\Phi}(k,r_a',r_b')~ j_l(kr_a')j_l(kr_b'). \een
We will next use the Limber approximation
Eq.(\ref{eq:limber_approx2}) to simplify the $k$ integral which
produces a $\delta_{1D}(r_a' -r_b')$ function. Integrating out
$r_b'$ with the help of the delta function and renaming the dummy
variable $r_a'$ to $r'$ we can finally write:

\ben
&& \myC_l^{\kappa\kappa}(k_1,k_2) = 
{2 \over \pi}k_1~k_2 \int_0^{\infty} r_1^2 dr_1 j_l(k_1r_1) \int_0^{\infty} r_2^2 dr_2 j_l(k_2r_2)
\myD_l^{\cred{\kappa\kappa}}(r_1,r_2); \nn \\
&& \qtwo  {\cred {\myD_l}}^{\kappa\kappa}(r_1,r_2) = { A^2 \over c^{\cred{4}}}
\int_0^{r_{min}} ~r'^2 {dr' \over a^2(r')}~ {F_\bbK(r_1,r')}{F_\bbK(r_2,r')}
P_{\delta} \left ( {l \over r'};r' \right ); \qquad r_{min} = min(r_1,r_2).
\label{kappa_cl}
\een
Use of the Limber approximation projects multi-time correlators to a single time correlator.
Going one step further, If we use the high $l$ approximation to the spherical Bessel function Eq.(\ref{eq:limber_approx3})
to reduce the dimensionality of the above integrals involving the spherical Bessel functions $j_l$, we arrive
at the following simpler approximate equation. Use of Eq.(\ref{eq:limber_approx3})
allows us to replace $r_1$ and $r_2$ in terms of $k_1,k_2$ and $l$.

\be
\myC_l^{\kappa\kappa}(k_1,k_2) = {[Ac^{-2}]}^2 \left [ 2 \over 2l + 1 \right ] \left [ {2l+1 \over 2k_1 }\right ]^2
\left [ {2l+1 \over 2k_2 }\right ]^2 \int_0^{r_{min}}
r'^2 {dr' \over a^2(r')}
F_\bbK\left [{2l+{1} \over 2k_1}, r'\right ]
F_\bbK\left [{ 2l+ {1} \over 2k_2},r' \right ]
P_{\delta}\left ( {l \over r'};r' \right).
\label{eq:cl21}
\ee

\n
We can define a statistic $\Sigma(k_1,k_2)$ which will include all available information from individual harmonics.
as a function of $k_1,k_2$:

\be
\Sigma^{\kappa\kappa}(k_1,k_2) = \sum_l (2l+1) \myC_l^{\kappa\kappa}(k_1,k_2); \quad
\Sigma^{\kappa\kappa}(r_1,r_2) = \sum_l (2l+1) \myC_l^{\kappa\kappa}(r_1,r_2).
\label{var}
\ee

\begin{figure}
\begin{center}
{\epsfxsize=6.cm \epsfysize=6.cm {\epsfbox[25 435 312 712]{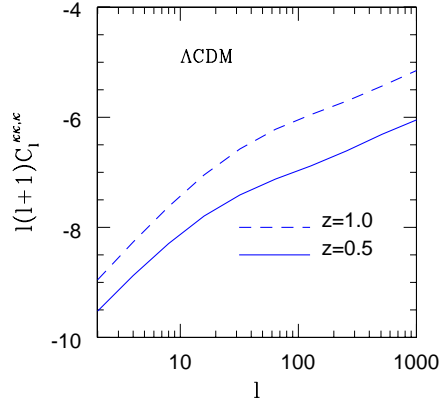}}}
\caption{The skew spectrum $C_l^{\kappa\kappa,\kappa}$ defined in Eq.(\ref{eq:cl21}) for the convergence $\kappa$ is plotted as a function
of $l$ for two different source redshifts $z_s=1$ (upper curve) and $z_s=0.5$ (lower curve).
A hierarchical form for the bispectrum was assumed
and the hierarchical amplitude is set to unity. As $\Lambda$CDM background cosmology is assumed.
We have not incorporated any smoothing window - calculations are done by introducing a sharp cutoff at $l_{max}=2000$.}
\label{fig:cl21}
\end{center}
\end{figure}

\n
We have ignored angular smoothing in our derivation. Typically observations will involve a
smoothing filter. Tophat and compensated filters are the ones that are most commonly used that can be
incorporated in $\label{eq:cl21}$. As pointed out before the above equation is derived using very general arguments.
It is valid at high $l$ as the derivation is based only on high $l$ approximation to the spherical Bessel function $j_l(x)$.
Nevertheless the derivation of a 3D skew spectrum has wider applicability in cosmology. The technique can be
applied to compute 3D power spectrum in other context (e.g. integrated Sachs-Wolfe effect or Kinetic Sunyaev-Zeldovich effect).
Detailed analysis for such cases will be presented elsewhere.

We have used Limber approximation to simplify results \citet{LoAf08}. It was pointed out by \citet{LoAf08}
that using $l$ instead of $l+{1 \over 2}$, as is often done in the literature, spoils the accuracy of Limber
approximation to ${\cal O}({1 \over l})$. In general the error in Limber approximation will scale as ${\cal O}({1 \over l^4})$.
A series expansion of spherical Bessel function can also be performed to construct the next to
leading order terms which further improve the accuracy of Limber approximation.

Weak lensing not only induces correlations among ellipticities of background galaxies (shear), it also introduces
local modification in the number density of source galaxies (also know as weak lensing magnification). The
weak lensing magnification $\mu$ is directly
linked to the convergence $\kappa$. While we have focused on shear $\gamma_{\pm}$ in this paper
we plan to extend the results to magnification in a related publication.

While the results derived above are valid for all-sky surveys, observations invariably will introduce
mask. We will next analyze the case of 3D power spectrum estimation in the presence of a general mask.
The results that we derive will have general applicability and will be valid for near all-sky coverage.

\subsubsection{The Effect of an angular sky Mask}

\n
We start by 3D decomposition of an arbitrary spinorial field ${\gg}$ in the presence of a sky mask $w(\oh)$, which is equal to 0 or 1 in simple cases.
The decomposition into radial harmonics, using spherical Bessel functions $j_l(kr)$, can be performed independent of the mask.
The harmonic decomposition on the surface of the celestial sphere
involves spin-weight spherical harmonics $_sY_{lm}(\oh)$. The following expression relates the masked observed harmonics
$[{\gg}w]_{lm}(k)$ and he unmasked $\gg_{lm}(k)$. We will leave the spinorial field ${\gg}$
arbitrary and derive the expression of the cross-correlation power spectrum, in the presence of a mask,
with another arbitrary spinorial field ${\ggp}(\oh)$. The results are generic and do not depend on any specific
assumption that is used to model the all-sky power spectrum itself.

\beqa
[\tilde {\gg}]_{lm}(k) \equiv &&
[{\gg}w]_{lm}(k) \equiv \sqrt{2 \over \pi}\int d^3\br kj_l(kr)[_sY^*_{lm}](\oh) [{\gg}(\br)w(\oh)]  =\sqrt{2 \over \pi}\int k r^2~dk j_l(kr)\int d\oh [{\gg}(\br)w(\oh)] [{}_{s}Y_{lm}^*(\oh)] \nn \\
&& = \sum_{l_bm_b} \left[ \sqrt{2 \over \pi}\int dk\; k\; r^2 \; j_l(kr)\gg_{l'm'}(r) \right ] w_{l_am_a}  \int [{}_{s}Y_{l'm'}(\oh)] [Y_{l_am_a}(\oh)]
[{}_{s}Y_{lm}^*(\oh)] d\oh \nn \\
&& = {2 \over \pi}\sum_{l'm'}\sum_{l_bm_b} (-1)^{s+m} \int dr\; \int dk' k\;k'\; r^2 \; j_l(kr)
 j_{l'}(k'r) \gg_{l_2m_2}(k') w_{l_am_a} J_{l'l_al} \left (  \begin{array} { c c c }
     l' & l_a & l \\
     s & 0 & -s
  \end{array} \right ) \left (\begin{array} { c c c }
     l' & l_a & l \\
     m' & m_a & -m
  \end{array} \right ); \\
&& \approx \sum_{l'm'}\sum_{l_am_a} (-1)^{s+m}{(2l'+1)^{1 \over 2} \over (2l+1)^{1 \over 2}} \gg_{lm}\left [ {2l+1 \over 2l'+1}k\right ]
w_{l_am_a} J_{l'l_al} \left (  \begin{array} { c c c }
     l' & l_a & l \\
     s & 0 & -s
  \end{array} \right ) \left (\begin{array} { c c c }
     l' & l_a & l \\
     m' & m_a & -m
  \end{array} \right ).
\label{individual_multipole}
\eeqa

\n
The masked cross-spectrum $\myC_l^{\sg\sgp}(k,k')$ involving $\sg(\oh)$ and $\sgp(\oh)$ can be described  in terms of the
all-sky cross-spectra $\myC_l^{\sg\sgp}(k,k')$ and a mode-mixing matrix $G$ that describes the effect of mode-mode
coupling resulting from the presence of the mask. The mode-mixing matrix depends on the spin-weight of the respective
fields $s$ and $s'$ and also depends on the power spectrum of the mask $w_{l}$.

\beqa
&& \tilde \myC_l^{\sg\sgp}(k,k') \equiv {1 \over 2l+1}\sum_{m}[\tilde{\gg}]_{lm}(k)[\tilde{\ggp}]_{lm}(k') \equiv {1\over 2l+1}\sum_{m}[{\gg}w]_{lm}(k)[{\ggp}w]_{lm}(k') \approx
\sum_{l'} G_{ll'} \myC_{l'}^{\sg\sgp}\left [ {2l+1 \over 2l'+1}k,{2l+1 \over 2l' + 1}k' \right ];\\
&& G_{ll'} = {1 \over 4 \pi}\sum_{l_a} {(2l'+1)^2\over(2l+1)} (2l_a+1)
\left (\begin{array} { c c c }
     l & l_a & l' \\
     s & 0 & -s
  \end{array} \right )
\left (\begin{array} { c c c }
     l & l_a & l' \\
     s' & 0 & -s'
  \end{array} \right )
|w_{l_a}|^2.
\label{mode_coupling_cls}
\eeqa

\n
The radial direction remains unaffected by the mask which introduces mode mixing only on the surface of the
celestial sphere hence the matrix $M$ is independent of radial wave number $k$. For a given pair of radial
wavenumbers $k,k'$ the all-sky cross-spectra
$C_l^{\sg\sgp}(k,k')$ can in principle be recovered by inverting the above expression Eq.(\ref{mode_coupling_cls}). In general
there will be contribution from noise which can be from intrinsic ellipticity or flexion distribution of galaxies in case of
shear or flexion. Such contributions need to be subtracted to make any estimation unbiased, and there may be standard issues with inversion which may require regularisation.

To recover the power spectrum of $E$ and $B$ modes of a spin $\pm2$
fields, commonly used in the context of analysis of CMB polarization
analysis, e.g. as in \citet{BCT},  we have to express angular
harmonics of $\gg=_{+2}\Gamma$ and $\ggp = _{-2}\Gamma$ in terms of
their Electric (E) and magnetic (B) components $_{2}\Gamma_{lm} =
E_{lm}+iB_{lm}$. Then using Eq.(\ref{mode_coupling_cls}) we can
relate the cut-sky  power spectra $\tilde \myC_l^{EE} = {1 \over
2l+1}\sum_m \tilde E_{lm}\tilde E^*_{lm}$ and $\tilde \myC_l^{BB} =
{1 \over 2l +1} \sum_m \tilde B_{lm}\tilde B^*_{lm}$ in terms of
their all-sky counterparts $\myC_l^{EE}$ and  $\myC_l^{BB}$. However
Eq.(\ref{mode_coupling_cls}) generalizes such results to generic
spin functions with arbitrary spin-weights. For generic spin weight
functions the cut-sky and all-sky relations are:

\be
\myC_l^{EE} = G^{EE}_{ll'}  \myC_{l'}^{EE} + G^{EB}_{ll'} \myC_{l'}^{BB}; \qquad\qquad
\myC_l^{BB} = G^{BE}_{ll'} \myC_{l'}^{EE} + G^{BB}_{ll'} \myC_{l'}^{BB}
\ee

\n
A sum over repeated indices are assumed in each of these equations.
The matrices $G^{EE}_{ll'}$, $G^{BB}_{ll'}$ and $G^{EB}_{ll'}$ are defined through the following expressions:

\beqa
&& G_{ll'}^{EE} = G_{ll'}^{BB} \equiv {1 \over 8\pi}\sum_{l_a} {(2l'+1)^2\over (2l+1)}(2l_a+1) (1+(-1)^L)\left (\begin{array} { c c c }
     l & l_a & l' \\
     s & 0 & -s
  \end{array} \right )^2 \nn \\
&& G_{ll'}^{EB} = G_{ll'}^{BE} \equiv {1 \over 8\pi}\sum_{l_a} {(2l'+1)^2 \over (2l+1)}(2l_a+1) ((-1)^L -1)\left (\begin{array} { c c c }
     l & l_a & l' \\
     s & 0 & -s
  \end{array} \right )^2;
\label{ps_EB}
\eeqa

\n
It is interesting to notice here that instead of $\myC_l(k_1,k_2)$ if we study
$\Sigma(k_1,k_2)=\sum_l(2l+1)\myC_l(k_1,k_2)$ they will
have exactly similar mixing properties as the ordinary 2D fields, modulo the remapping of the radial harmonics,
as the usual 3D power spectrum $C_l$ when a mask is applied, i.e.
${\tilde \Sigma}^{\kappa\kappa}(k_1,k_2) = \sum_{l'} M_{ll'} \Sigma_{}({2l+1 \over 2l' + 1}k_1,{2l+1 \over 2l' + 1}k_2) $;
where $G_{ll'} = {(2l'+1)\over(2l+1)}M_{ll'}$.
This property will be valid not just as the level of power spectrum but also for skew- and kurt spectra
as well as for multispectra of arbitrary order.

We have plotted the projected power spectrum for convergence $\myC_l^{\kappa\kappa}$ in Fig. (\ref{fig:cl_ps}) as
a function of $l$ (left panel). Two different redshifts were considered $z_s=0.5$ and $z_s=1.0$.
The $\Lambda$CDM background cosmology that we will be using throughout this paper is characterised
by the following set of parameters: $\Omega_m=0.3$, $\Omega_\Lambda=0.7$, $\Gamma=0.21$, $h=0.7$ and $\sigma_8=0.90$.
The power spectra associated with other spinorial fields (shear and flexions) are also plotted (right panel).
The 3D power spectrum
$\myC_l^{\kappa\kappa,\kappa}$ is plotted in Fig. \ref{fig:clkk} for the same background cosmology. We plot
$\myC_l^{\kappa\kappa,\kappa}(k,k)$ for three different choice of $k$ values as function of $l$ (left panel) as
well as $C_l^{\kappa\kappa,\kappa}(k,k)$ for three selection of $l$ values as a function of the radial wave number $k$.

\subsection{Bispectrum}

The power spectrum carries the bulk of the information in any cosmological observations. However often
a set of degenerate cosmological scenarios can lead to a very similar power spectrum. Analysing
higher-order correlation functions can lift this degeneracy to some extent.
The non-Gaussianity used can be either due to primordial or secondary effects, and in the case of weak lensing the
main source of non-Gaussianity comes from gravitational instability. Note that a non-zero bispectrum signifies the lowest-order departure from gaussianity,
and its detection is generally easier than higher-order multispectra.

To make contact with the observables we use the fact that the convergence can be related directly to the 3D density field. We will start by linking the
3D convergence bispectrum ${\cal B}$ and the 3D density bispectrum expressed in harmonic coordinates. In the next
section we will express the bispectrum in spherical coordinate in terms of the bispectrum in rectangular coordinates
and use some well-motivated approximations to simplify the results.

Statistical isotropy requires that
\begin{equation}
\cg{\langle \kappa_{l_1 m_1}(k_1;r_1) \kappa_{l_2 m_2}(k_2;r_2) \kappa_{l_3 m_3}(k_3;r_3)\rangle = \left(
  \begin{array}{ c c c }
     l_1 & l_2 & l_3 \\
     m_1 & m_2 & m_3
  \end{array} \right)
{\cal B}_{l_1l_2l_3}^{\kappa\kappa\kappa}(k_i;r_i)}
\label{bi_def}
\end{equation}
and using Eq.(\ref{kappa2den}) we can write
\ben && {\cal B}^{\kappa\kappa\kappa}_{l_1l_2l_3}(k_i;r_i) = A^3 \cred{{\cal L}_1{\cal L}_2{\cal L}_3} \left ( {\cred{2} k_1
\over \pi c^2 }\right ) \left ( {\cred{2} k_2 \over \pi c^2 }\right ) \left
( {\cred{2} k_3 \over \pi c^2 }\right )
\inte  { dk_1' \over k_1'} \inte dr_1 r_1^2 j_{l_{\cred{1}}}(k_1'r_1')\cred{j_{l_1}(k_1r_1)} \int_0^{r_1} {dr_1' \over a(r_1')} F_K(r_1,r_1')\times \nn \\
&& \inte {dk_2' \over k_2'} \inte dr_2 r_2^2 j_{l_2}(k_2'r_2')\cred{j_{l_2}(k_2r_2)}
\int_0^{r_2} {dr_2' \over a(r_2')} F_K(r_2,r_2') \inte {dk_3' \over k_3'} \inte dr_3^2 r_3 j_{l_{\cred{3}}}(k_3'r_3') \cred{j_{l_3}(k_3r_3)}
\int_0^{r_3} {dr_3' \over a(r_3')}F_K(r_3,r_3')B^{\delta}_{l_1l_2l_3}(k_i';r_i'); \nn \\
\label{eq:con_bi}
&& \qquad\qquad\qquad\qquad\qquad\qquad\qquad\qquad~~ \cred{{\cal L}_i = l_i(l_i+1) \sim l_i^2}. \een

\n
The  bispectrum ${\cal B}^{\kappa\kappa\kappa}_{l_1l_2l_3}(k_i;r_i)$ can now
expressed in terms of the underlying matter bispectrum $B^{\delta}$.
The above relation mixes modes only in the radial
directions $r$, and on the surface of the sky there is no mixing of angular
harmonics if there is no sky mask. While expressing the
density harmonics in terms of the 3D potential harmonics, we pick up
additional scale factor $a(r_i)$ and wavenumber $k_i$ dependence in the
denominator.

We have so far ignored the presence of noise. Indeed because of the limited number of galaxies
available it may not be possible to probe individual modes of
the bispectrum at high signal-to-noise ratio. In later sections we will be able to address issues related to optimum combinations
of individual modes which may be better suited for observational studies.

The convergence bispectrum can be written in terms of the density bispectrum as follows, using the Limber approximation to simplify the results:

\ben && {\cal B}^{\kappa\kappa\kappa}_{l_1l_2l_3}(k_i;r_i) =
\cred{H_1H_2H_3} \inte r_1^2 dr_1 j_{l_1}(k_1r_1) \inte r_2^2 dr_2 j_{l_2}(k_2r_2) \inte r_3^2 dr_3
j_{l_3}(k_3r_3)
{\cal I}^{\cred{(3)}}_{l_1l_2l_3}(r_1,r_2,r_3); \quad \cred{H_i \equiv A~{k_i \over c^2}}}
 {\cred{\sqrt {2 \over \pi}} \nn \\
&&{\cal I}^{\cred{(3)}}_{l_1l_2l_3}(r_1,r_2,r_3) \equiv {\cred {S^{}}}_{l_1l_2l_3}b_{l_1l_2l_3}= {\cred {S^{}}}_{l_1l_2l_3}
\int_0^{r_{min}} r^2 dr
B^{\delta}\left ({l_1\over r},{l_2\over r},{l_3\over r}; r,r,r \right )
~{\cal R}_1(r) {\cal R}_2(r) {\cal R}_3(r); \quad \cg{{\cal R}_i(r) =  {F_K(r_i,r) \over a(r)}}.\nn \\
&& S_{l_1l_2l_3} = \sqrt{(2l_1 + 1)(2l_2+1)(2l_3+1)\over 4\pi}\left(
  \begin{array}{ c c c }
     l_1 & l_2 & l_3 \\
     0 & 0 & 0
  \end{array} \right).
\label{ConB}
\een

\cred{To derive this result we have used the extended Limber approximation Eq.(\ref{eq:limber_approx2}) to simplify the
$k_i'$ integrals.} The integral here extends to the overlapping region i.e. $r_{min} = min(r_1,r_2,r_3)$.
In particular we can use the gravity-induced bispectrum here, or include others, such as a primordial bispectrum.

It is possible to simplify  further the above expression using Eq.(\ref{eq:limber_approx1}):

\be
{\cal B}_{l_1l_2l_3}^{\kappa\kappa\kappa}[k_1,k_2,k_3] = \left [ Ac^{-2}\right ]^3
 \left [ {1 \over k_1k_2k_3} \right]^{2}
 \left [{2l_1 +1 \over 2} \right ]^{3/2} \left [{2l_2 + 1 \over 2} \right ]^{3/2} \left [ {2l_3 + 1 \over 2} \right ]^{3/2}
S_{l_1l_2l}
{\cal I}^{\cred{(3)}}_{l_1l_2l_3}\left ({2l_1 +1 \over 2k_1},{2l_2 + 1 \over 2k_2},{2l_3 + 1 \over 2k_3} \right ).
\ee
We will use a generic hierarchical ansatz to model the matter correlation hierarchy. Such ansatze
constructs multi-point correlation functions from the products of lower-order correlation functions.
In the Fourier domain this will lead to the construction of multispectra from products of ordinary power spectra.
Such models have been tested against simulations and are routinely used both for projected
galaxy surveys and for 2D weak lensing surveys. At the level of bispectrum we have: $B(k_1,k_2,k_3,r) = Q_3[P(k_1,r)P(k_2,r) +
P(k_2,r)P(k_3,r) + P(k_1,r)P(k_3,r)]$. More detailed modelling will make $Q_3$ a function of the wave vector triplet ($k_1,k_2,k_3$) e.g. in the
halo model \cite{CooSeth02} or in Hyper Extended Perturbation Theory \citep{Scocci98}. While the
expression derived above is for the convergence field $\kappa$,
it can be used to construct the other bispectra involving shear or flexion:

\be
{\cal B}^{\gamma_{\pm}\kappa\kappa}_{l_1l_2l_3}(k_i) = F^{\gamma_{\pm}}_{l_1} {\cal B}^{\kappa\kappa\kappa}_{l_1l_2l_3}(k_i); \quad
{\cal B}^{\gamma_{\pm}\gamma_{\pm}\kappa}_{l_1l_2l_3}(k_i) = F^{\gamma_{\pm}}_{l_1}F^{\gamma_{\pm}}_{l_2} {\cal B}^{\kappa\kappa\kappa}_{l_1l_2l_3}(k_i); \quad
{\cal B}^{\gamma_{\pm}\gamma_{\pm}\gamma_{\pm}}_{l_1l_2l_3}(k_i) = F^{\gamma_{\pm}}_{l_1}F^{\gamma_{\pm}}_{l_2}F^{\gamma_{\pm}}_{l_3} {\cal B}^{\kappa\kappa\kappa}_{l_1l_2l_3}(k_i); \quad
\label{individual_bispec}
\ee
Results involving flexion can be constructed replacing the form factor $F_l^{\gamma_{\pm}}$s with the ones
for the flexions i.e. $F_l^{\cal F}$ or  $F_l^{\cal G}$ defined accordingly. The radial dependence of
convergence, shear or flexion harmonics are the same.

Mode coupling is introduced by the presence of an observational mask. It is not possible to deconvolve
the effect of a mask while analyzing the bispectrum from a realistic survey, as the inversion is typically unstable. However later we will
introduce a power spectrum associated with the bispectrum (the skew spectrum), which can be computed from realistic data in the presence of mask
and deconvolution can be done in a way very similar to the estimation of power spectrum discussed
previously.

If we assume the intrinsic shear and flexion of source galaxies to be distributed
according to a Gaussian distribution, then they do not contribute to the estimated bispectrum, but this is, of course, an assumption.
However even in this case the scatter or variance in estimation does get a contribution from such a source of noise.

It is important to realize the results derived above are generic. They do not depend on
the specific model used as an example (hierarchical ansatz). If we replace the underlying
bispectrum with a primordial bispectrum of a specific type (e.g. local) we can still use the
formalism developed here to compute various relevant statistics, we will introduce later, e.g. the skew spectrum.
Later we will also introduce an optimized estimator for the skew spectrum. This estimator is not only
optimized to detect any specific type of non-Gaussianity but it can also give an estimate
of leakage from a specific source of non-Gaussianity (e.g. gravity-induced) while estimating another (primordial).

\subsection{Trispectrum}

The trispectrum or, alternatively, the four-point correlation
function can provide an important sanity check  \citep{KSH10} to
validate lower-order detection of non-Gaussianity based solely on
the bispectrum. The trispectrum, being a four-point correlation
function is generally harder to probe compared to the bispectrum.
However, in cases where the bispectrum vanishes due to symmetry
considerations, the trispectrum is the lowest probe to study
gravity-induced non-Gaussianity. In addition,  while the study of
trispectrum is interesting in itself it is also important in the
proper characterization of the error in the power spectrum. As in
the case of bispectrum we will start by modelling the bispectrum of
the convergence field which can then be generalised to model
trispectra associated with various spinorial fields. These results
will eventually be used to model two different power spectra
associated with the trispectrum (the {\it kurt spectra}).

The convergence trispectrum ${\cal T}^{l_1l_2}_{l_3l_4}(L,k_i;r_i)$ is the four-point correlation function
in the harmonic domain and can be expressed as

\begin{equation}
\langle \delta_{l_1m_1}(k_1;r_1)\delta_{l_2m_2}(k_2;r_2)\delta_{l_3m_3}(k_3;r_3)\delta_{l_4m_4}(k_4;r_4)\rangle_c  = \sum_{LM} (-1)^M
  \left( \begin{array}{ c c c }
     l_1 & l_2 & L \\
     m_1 & m_2 & M
  \end{array} \right)
  \left(  \begin{array}{ c c c }
     l_3 & l_4 & L \\
     m_3 & m_4 & -M
  \end{array} \right) {}^\delta{T}^{l_1l_2}_{l_3l_4}(L,k_i;r_i).
\label{eq:tri}
\end{equation}
The vectors $l_1,l_2,l_3,l_4$ represents the sides of a quadrilateral and L is the length of the diagonal.
The matrices as before are the Wigner $3j$ symbols. The symbols are only non-zero when they satisfy
several conditions; which are $|l_1-l_2| \le L \le l_1+l_2$, $|l_3-l_4| \le L \le l_3+l_4$;
$l_1+l_2+L$= even, $l_3+l_4+L$ = even and $m_1 + m_2 = M$ as well as $m_3+m_4 = -M$.
In our notation for the trispectrum, $T^{l_1l_2}_{l_3l_4}(k_i,r_i;L)$, the indices $(k_i\cg{;}r_i)$
encode their dependence on various Fourier modes of the density harmonics in the radial direction, used
in their construction. No summation will assumed over these variables
unless explicitly specified. To model the trispectrum we will relate it to the underlying trispectrum of
the density distribution $T^{l_1l_2}_{l_3l_4}(k_i';r_i')$.

\ben && {}^{\kappa}{\cal T}^{l_1l_2}_{l_3l_4}(k_i;r_i) =  A^4 {\cal L}_1{\cal L}_2{\cal L}_3{\cal L}_4
\left ({2 k_1  \over \pi c^2 }\right )\cdots \left ({2 k_4  \over \pi c^2 }\right )  \inte  { dk_1' \over k_1'} \inte dr_1 r_1^2 j_{l_1}(k_1'r_1') \int_0^{r_1} {dr_1' \over a(r_1')} F_K(r_1,r_1') \nn \\
&& \qquad \qquad \qquad \qquad \qquad \qquad   \dots \inte {dk_4'
\over k_4'} \inte dr_4 r_4^2 j_{l_4}(k_4'r_4') \int_0^{r_4} {dr_4'
\over a(r_4')} F_K(r_4,r_4')\;\;
{}^{\delta}T^{l_1l_2}_{l_3l_4}(k_i';r_i'). \een

\n
The above expression is a direct consequence of Eq.(\ref{Den}).
To make any further progress we need to consider a specific form for the matter trispecrum.
There are two generic prescriptions for treating a gravity-induced trispectrum.
The halo model is one, and has been developed over the last several years and is very popular
for modelling the correlation hierarchy of the underlying mass-distribution. Another is that
perturbative descriptions can also provide a
reasonable description of the onset of non-linearity at comparatively larger length scales.
The hierarchical ansatz on the other hand describes the matter correlation hierarchy
in the highly non-linear regime on smaller length scales. It builds up the higher-order correlation
hierarchy from the two-point correlation function. All possible {\it diagrams} that connect points
at which the correlation function is being constructed are considered. These
diagrams are attributed various amplitudes according to their topology.
Different diagrams with same topologies are associated with same amplitude.
At the level of the four-point correlation function there are only two different topologies {\it star} and {\it snake}.
We will denote the corresponding amplitudes by $R_b$ and $R_a$ and consider
each of these contributions separately next.

\subsubsection{Star Diagrams}

The star diagrams are easier to handle because topologically they
consist of a single vertex and lack any internal momentum that needs
to be integrated out. In generic hierarchical scenarios a {\it star}
diagram appears at each order in the hierarchy. In the case of
projected 2D analysis, it has been found that replacing all diagrams
with the same number of star diagrams often is sufficient to
reproduce all one-point statistical features of convergence maps
\citep{MuVaBa04,BaMuVa04,VaMuBa05}. However the results presented
here are generic and includes both contributions.

The derivation of the star contribution to the convergence trispectrum follows
a similar technique as the bispectrum. The first step is to
express the density convergence in terms of the triplets of matter power spectra .
Next, using the expression Eq.(\ref{Den}) we can relate the star contribution of convergence
trispectra to that of underlying mass distribution:
\ben
&& {}^{\delta}T^{l_1l_2}_{l_3l_4} \left (L,{l_i \over r}; r_i \right )_{\rm star} =
\left ( {2 \over \pi} \right )^2 k_1k_2k_3k_4 \int_0^{\infty} dr_1 r_1j_{l_1}(k_1r_1) \dots \int_0^{\infty} dr_4 r_4j_{l_4}(k_4r_4) J^{(4)}_{l_1l_2l_3l_4}(r_1,r_2,r_3,r_4)_{\rm star},
\quad {\rm where} \nn \\
&& {J}^{\cred{(4)}}_{l_1l_2l_3l_4}(\cg{k_i;r_i})_{\rm star} \equiv \cred{R_b}
\cred{S}_{l_1l_2L}\cred{S}_{l_3l_4L} \int r^2 dr j_{l_1}(k_1r) \dots j_{l_4}(k_4r)\left \{ P(k_1;r_1)P(k_2;r_2)P(k_3;r_3) +
{\rm cyc.perm.} \right \}.
\een

The above equation is for the stellar contribution to the
trispectrum of the underlying density distribution. The additional
three terms can be recovered by cyclic permutation of the $k_i$
variables. We can use this result next to express the convergence
trispectrum. The simplification relies on the use of the extended
Limber approximation to simplify the $k_i$ integrals. Hierarchical
ansatz and Limber approximations are both known to be valid at small
scales which justifies their combined use \citep{MuHeCo_wl1_10}.

\ben
&& {}^{\kappa}{\cal T}^{l_1l_2}_{l_3l_4} \left (L,{l_i \over r}; r_i \right )_{\rm star} =
\cred{H_1H_2H_3H_4} \inte r_1^2 dr_1 j_{l_1}(k_1r_1) \cdots 
\inte r_4^2 dr_4 j_{l_4}(k_4r_4)
{\cal I}^{\cred{(4)}}_{l_1l_2l_3l_4}(r_1,r_2,r_3 \cred{,r_4})_{\rm star} \qquad {\rm where} \nn \\
&& {\cal I}^{(4)}_{l_1l_2l_3\cred {l_4}}(r_1,r_2,r_3 \cred{,r_4})_{\rm star} = 
\cred{R_b} \cred{S}_{l_1l_2L}\cred{S}_{l_3l_4L} \int_0^{r_{min}} r^2 dr  {\cal R}_1(r) \dots {\cal R}_4(r)
\left \{ P \left ({l_1\over r};r \right )P \left ({l_2\over r};r \right )P \left ({l_3\over r};r \right ) + \rm{cyc. perm.} \right \}.
\label{star}
\een

\n
In generic hierarchical scenarios the trispectrum is a  cubic combination of underlying matter power spectra,
just as the convergence bispectrum is a quadratic function of matter power spectra. We will focus on
a specific model, the hierarchical ansatz, to model the underlying density distribution. However as it was
pointed out that similar techniques can also be applied in the context of more elaborate halo model
prescription.

\subsubsection{{Snake Diagrams}}

The analysis of the snake diagrams is more difficult than that of
the star diagrams. This is related to the fact that while
higher-order {\it star} diagrams at each order are straightforward
generalizations of the lower-order star diagrams, the {\it snake}
diagrams are however constructed using two different lower-order
star diagrams. The following expression was derived in
\cite{MuHeCo_wl1_10} which relates the {\it star} contribution to
convergence trispectra.  In addition to various form factors, the
following expression depends on the cubic product of the underlying
matter power spectrum. The relative importance of star and snake
differs in different models of the hierarchical ansatz, e.g.
\cite{SzaSza93,SzaSza97} attributes equal weighting to both
diagrams, whereas \cite{BerSch92} constructs higher-order amplitudes
from the lower-order star amplitudes. The amplitude of the new star
diagram at each order is left arbitrary which can be fixed using
simulations or through the use of hyper-extended-perturbation theory
\citep{HEPT} which predicts one-point cumulants at each order. We
note in passing that ordinary perturbation theory predicts a
hierarchy similar to hierarchical ansatz at the onset of
gravitational clustering. However the amplitudes in this regime for
various topologies are different \citep{Fry84}.

\beqa
&& {\cal Q}^{l_1l_2}_{l_3l_4} \left (L,{l_i \over r}; r_i \right )^{\rm sph}_{\rm snake} =
{H_1H_2H_3H_4} \inte r_1^2 dr_1 j_{l_1}(k_1r_1) \cdots \inte r_4^2 dr_4 j_{l_4}(k_4r_4)
{\cal I}^{{(4)}}_{l_1l_2l_3l_4}(r_1,r_2,r_3 {,r_4})_{\rm snake} \nn \\
&& {\cal I}^{(4)}_{l_1l_2l_3l_4}(r_1,r_2,r_3,r_4)_{\rm snake} =  R_a S_{l_1l_2L}S_{l_3l_4L}
\int_0^{r_{min}} r^2 dr  {\cal R}_1(r) \dots {\cal R}_4(r) \left \{  P\left ({l_1 \over r},r\right )
P\left ({l_3 \over r},r \right )P\left ({L \over r},r \right )
+ {\rm cyc. perm.} \right \}.
\label{snake}
\eeqa

\n
The cyc. perm. here represents a total of three other terms. These terms can be obtained by rearranging $l_i$s
$(l_1 \rightarrow l_2)$, $(l_3 \rightarrow l_4)$ and $(l_1 \rightarrow l_2, l_3 \rightarrow l_4)$.
The other terms that can obtained by considering two additional pairings by considering the
exchanges  $(l_2 \rightarrow l_3)$ and  $(l_2 \rightarrow l_4)$. These will lead us to
${\cal Q}^{l_1l_3}_{l_2l_4}$ and ${\cal Q}^{l_1l_4}_{l_3l_2}$. The total number of {\it snake} terms considering
three distinct pairings and permutations within each pairings is twelve. The snake contribution to trispectrum
can be written as:

\beqa
{\cal T}^{l_1l_2}_{l_3l_4} \left (L,{l_i \over r}; r_i \right )^{\rm sph}_{\rm snake} = &&
 {\cal Q}^{l_1l_2}_{l_3l_4} \left (L,{l_i \over r}; r_i \right )^{\rm sph}_{\rm snake} +
(2L+1)\sum_{L'} (-1)^{l_2+l_3} \left\{  \begin{array}{ c c c }
     l_1 & l_2 & L \\
     l_4 & l_4 & L'
  \end{array} \right \}{\cal Q}^{l_1l_3}_{l_2l_4} \left (L',{l_i \over r}; r_i \right )^{\rm sph}_{\rm snake} \nn \\+
&& (2L+1)\sum_{L''}(-1)^{L+L''} \left \{  \begin{array}{ c c c }
     l_1 & l_2 & L \\
     l_3 & l_4 & L''
  \end{array} \right \}
{\cal Q}^{l_1l_4}_{l_3l_2} \left (L'',{l_i \over r}; r_i \right )^{\rm sph}_{\rm snake}
\label{6j}
\eeqa

\begin{figure}
\begin{center}
{\epsfxsize=12.cm \epsfysize=6.cm {\epsfbox[28 426 590 712]{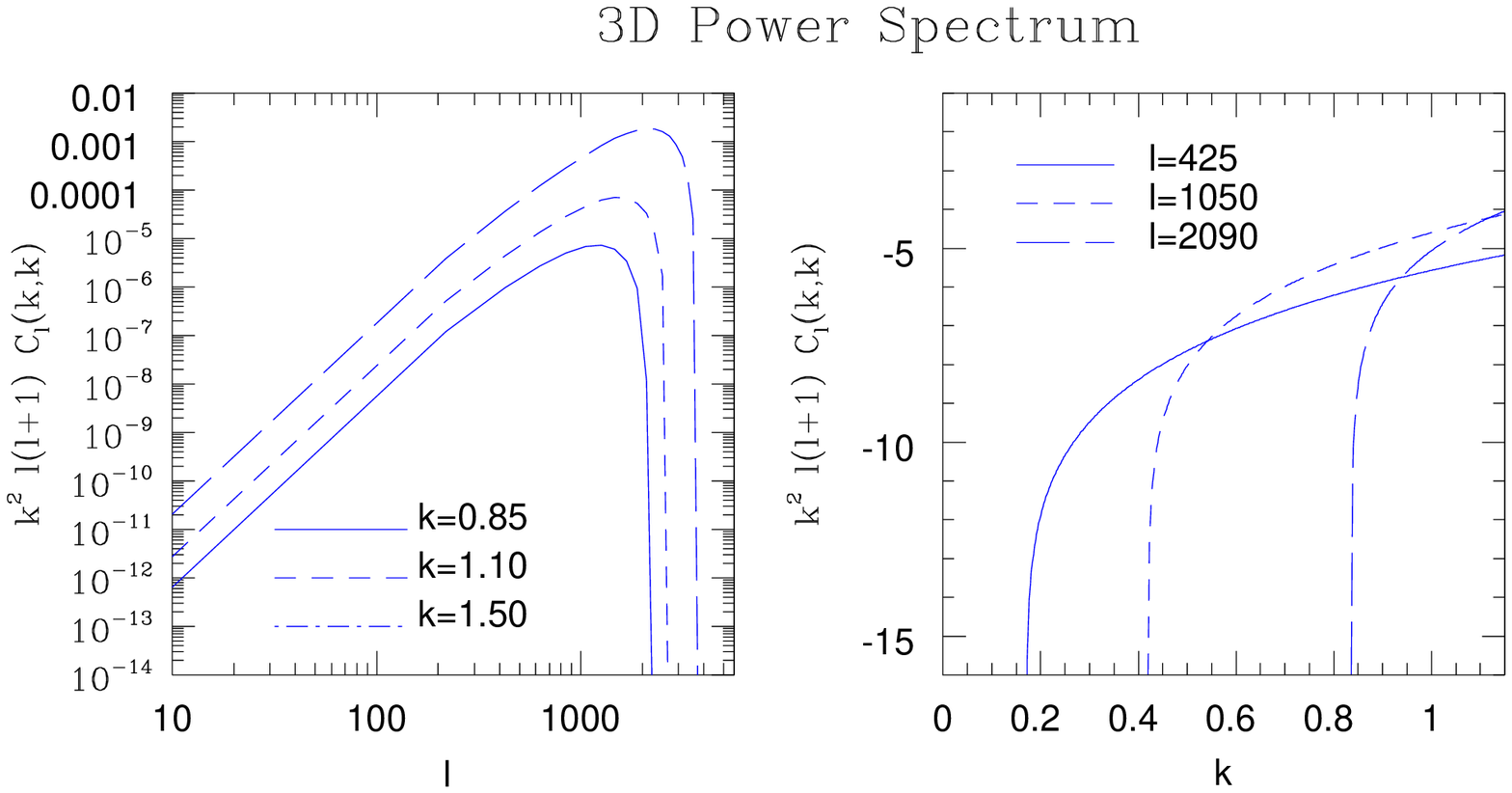}}}
\caption{The left panel shows 3D power spectrum $l(l+1)k^2C_l^{\kappa\kappa}(k,k)$ as a function of $k$ for three different
values of $l$. Where as the right panel shows $l(l+1)k^2C_l^{\kappa\kappa}(k,k)$ as a function of $k$ for fixed value of $l$.
A LCDM background cosmology was assumed with $\Omega_m=0.3$,
$\Omega_{\Lambda}=0.7$, $h=0.7$, $\sigma_8=0.8$, $n_s=1.0$, $w=-1$; we use an Eisenstein \&
Hu (1996) linear power spectrum and the Smith et al. (2003) non-linear correction. The
sharp cutoff in these plots reflects survey depth through the Bessel inequality. Here $k$ is displayed in ${\rm hMpc}^{-1}$.}
\label{fig:clkk}
\end{center}
\end{figure}
The matrices in curly braces are the $6j$ symbols \citep{Ed68}. The
differences in {\em snake} and {\em star} contributions are also
apparent in various choices of permutations of $(l_1,l_2,l_3,l_4;L)$
associated with individual terms.

The derivations outlined both for bispectrum and trispectrum are quite generic and depend
only on the use of extended Limber approximation, and the analysis can be
generalized to higher-order multispectra.

\n
We will use these results to construct the power spectra associated with the trispectrum,
or kurt spectra. Construction of trispectra for generic spin-weight functions
can be achieved by using the form factors in a manner similar to the one adopted for the
bispectrum Eq.(\ref{individual_bispec}).

\be
{\cal T}^{\sg_{l_1}\sgp_{l_2}}_{\sgpp_{l_3}\sgppp_{l_4}}(L,k_i) =
F^{\sg}_{l_1}F^{\sgp}_{l_2}F^{\sgpp}_{l_3}F^{\sgppp}_{l_4}{\cal T}^{l_1l_2}_{l_3l_4}(L,k_i)
\ee

\n
The two different kurt spectra  that we will construct next will provide independent probes of the underlying mass
trispectra, and can play a valuable role in checking any cross-contamination from systematics.

\section{Power Spectra Associated with Multispectra}

The multispectrum may encode a great deal of information, but there
is certain amount of degeneracy involved in it. Owing to the low
signal--to --noise associated with the estimation of multispectra,
it is impossible to estimate  multispectra mode by mode. Estimation
of multispectra is also hampered by their complicated response to
the survey mask and complicated noise characteristics. Recent
studies have shown that degenerate sets of power spectra can be
constructed from multispectra at a given order. These compress some
of the available information in the multispectra and can be computed
from the observational data with relatively ease. We will construct
the power spectra associated with bispectrum and trispectrum in this
section. These power spectra were studied in some detail in
\citep{Mu_wl2_10} in projection (2D). We generalize these results
here to 3D. First we will obtain generic results, applicable
irrespective of detailed analytical modelling of underlying
multispectra.Next we will further specialize these results for the
models outlined above, and ee use extended Limber approximation to
simplify the generic results.

\subsection{skew spectrum}

\subsubsection{All-Sky Results}
\n
We will start by expanding the product of two generic spinorial fields with associated spin indices $s,s'$
$[{\gg}{\ggp}]({\bf r})$ in 3D in terms of their
individual harmonics. The product of two spinorial fields with spins $s$ and $s'$ is a spinorial field of spin $s+s'$.
The harmonic expansion therefore will have to be in terms of spin-weight spherical harmonics with spin index $s+s'$.
In addition to expanding on the surface of the celestial sphere,
we will also consider the expansion in the radial direction using spherical Bessel functions.

\begin{eqnarray}
&& [{\gg}({\bf r}){\ggp}({\bf r})]_{lm}(k) = \sqrt{ 2 \over \pi}\int_0^{\infty} k r^2 dr j_l(r k) \int ~d\oh ~{\gg}({\bf r})~{\ggp}({\bf r})[{}_{s+s'}Y_{lm}^*(\oh)] \nn \\
&& ={\left (2 \over \pi \right )}^{3/2} \int_0^{\infty} r^2 dr k j_l(r k)
\sum_{l_im_i} \int_0^{\infty} k_1 dk_1 j_l(k_1r)\gg_{l_1m_1}(k_1)\int_0^{\infty} k_2 dk_2 j_l(k_2r)\ggp_{l_2m_2}(k_2)
\int [{}_sY_{l_1m_1}(\oh)]~[{}_{s'}Y_{l_2m_2}(\oh)]~
[{}_{s+s'}Y_{lm}^*(\oh)] d\oh \nn \\
&&  = {\left ( 2 \over \pi \right )}^{3/2}\sum_{l_im_i}\int_0^{\infty} dr k r^2 dr j_l(kr) \int_0^{\infty} k_1 dk_1 j_l(k_1r) \gg_{l_1m_1}(k_1)
\int_0^{\infty} k_2 dk_2 j_l(k_2r)\ggp_{l_2m_2}(k_2) \nn \\
&& \quad\quad \times J_{l_1l_2l} 
\left ( \begin{array} { c c c }
     l_1 & l_2 & l \\
     s & s' & -(s+s')
  \end{array} \right )\left ( \begin{array} { c c c }
     l_1 & l_2 & l \\
     m_1 & m_2 & m
  \end{array} \right ); \quad J_{l_1l_2l} = \sqrt{(2l_1+1)(2l_2+1)(2l+1) \over 4 \pi}.
\label{eq:decompose}
\end{eqnarray}

\n
We have used the Gaunt integral to express integrals involving spherical harmonics
in terms of the Wigner $3j$ symbols at the last step. The matrices describe the Wigner 3j symbols.
To construct the skew spectrum, next we contract the multipole
$[{\gg}({\bf r}){\ggp}({\bf r'})]_{lm}$ with the multipole associated with a third spinorial
3D field  $\ggpp^*_{lm}(k_3)$. The resulting skew spectrum or power spectrum associated
with the bispectrum depends on the three radial wave numbers $(k_1,k_2,k_3)$, in addition to
the spinorial indices associated with the respective 3D fields:

\be C_l^{\sg \sgp, \sgpp}(k,k') \equiv  {1 \over 2l+1} \sum_m
[\gg\ggp]_{lm}(k) \ggpp^*_{lm}(k') \label{eq:bispec} \ee The
bispectrum is defined by a triangular configuration in multipole
space with lengths of side $(l_1,l_2,l)$. The power spectrum
constructed above, will essentially capture information, through a
summation, with all triangular configurations with one of these
sides kept fixed at length $l$. If we now use the following
expression for the 3D bispectrum
$B_{l_1l_2l_3}^{\sg\sgp\sgpp}(k_1,k_2,k_3)$ introduced earlier,

\be
{\cal B}_{l_1l_2l_3}^{\sg\sgp\sgpp}(k_1,k_2,k_3;r_i) = \sum_{m_1m_2m_3} \langle\gg_{l_1m_1}(k_1;r_1) \ggp_{l_2m_2}(k_2;r_2)
\ggpp_{l_3m_3}(k_3;r_3) \rangle_c
\left ( \begin{array} { c c c }
     l_1 & l_2 & l_3 \\
     m_1 & m_2 & m_3
  \end{array} \right ),
\label{gen_bi}
\ee
\begin{figure}
\begin{center}
{\epsfxsize=12.cm \epsfysize=6.cm {\epsfbox[28 426 575 712]{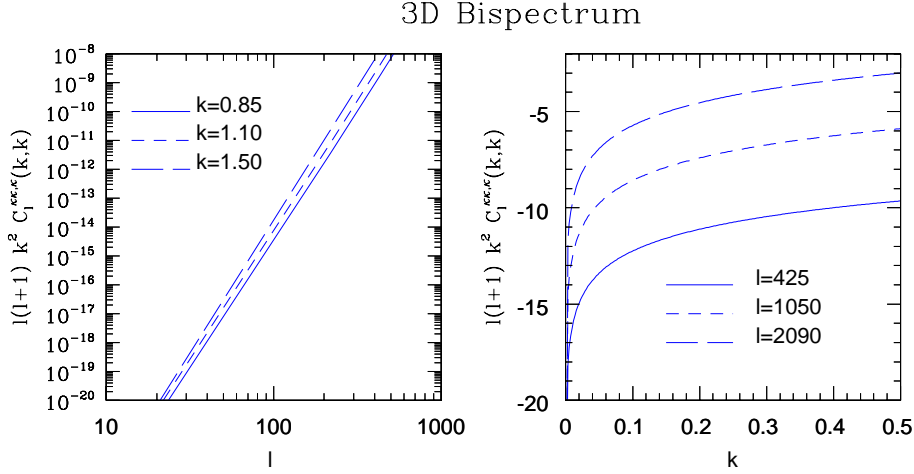}}}
\caption{The right panel shows 3D power spectrum $l(l+1)k^2C_l^{\kappa\kappa,\kappa}(k,k)$ as a function of $k$ 
for three different
values of $l$. Where as the left panel shows $l(l+1)k_2^2C_l^{\kappa\kappa,\kappa}(k,k)$ as a function of $l$ for fixed value of $k$.
Three different $k$ values were chosen as depicted. 
A $\Lambda$CDM background cosmology was assumed with $\Omega_m=0.3$,
$\Omega_{\Lambda}=0.7$, $h=0.7$, $\sigma_8=0.8$, $n_s=1.0$; we use an Eisenstein \&
Hu (1996) linear power spectrum and the Smith et al. (2003) non-linear correction.
A hierarchical ansatz was assumed for the bispectrum with the amplitude fixed at $Q=Q_3=1$. See text for more details. 
Here the wave vector $k$ is displayed in units of ${\rm hMpc}^{-1}$.}
\label{fig:cl21kk}
\end{center}
\end{figure}

we can express the two-to-one  power spectrum as:
\beqa
 \myC_l^{\sg\sgp,\sgpp}(k,k') = \left ( {2 \over \pi} \right ) \int_0^\infty ~r^2 dr~ k j_l(kr)
\sum_{l_1l_2} J_{l_1l_2l}
\left ( \begin{array} { c c c }
     l_1 & l_2 & l \\
     s & s' & -(s+s')
  \end{array} \right )
\int_0^{\infty} k_1 dk_1 j_l(k_1r) \int_0^{\infty} k_2 dk_2 j_l(k_2r) {\cal B}_{l_1l_2l}^{\sg\sgp\sgpp}(k_1,k_2,k';r,r,r').
\label{exact_bicls}
\eeqa

\n Although the above relation is general and contains all the
information regarding the evaluation of the skew spectrum for
arbitrary spin-weight functions, the presence of multidimensional
integrals makes it difficult to implement computationally.

We will simplify the results by using the extended Limber
approximation to reduce the integrals involving $k_1$ and $k_2$.
First we express the $\cal B$ in terms of the underlying matter
bispectrum  $\rm B$ using the expression Eq.($\ref{ConB}$). The
extended Limber approximation, Eq.\;(\ref{eq:limber_approx2}),
collapses the $k_i$ integrals to one-dimensional Dirac delta
functions. These delta functions are next used to reduce the $r_1$
and $r_2$ integrals. These simplifications which are known to be
valid at high $l$. Finally we are left with two integrals involving
$r$ and $r_3$ which can be further simplified by
Eq.\;(\ref{eq:limber_approx3}). The final expression contains only a
one-dimensional integral and directly relates the skew spectrum
$\myC_l^{\sg\sgp,\sgpp}(k,k')$ to the matter bispectrum:

\be
C_l^{\sg\sgp,\sgpp}(k,k') = \sum_{l_1l_2} \left [ {2 \over 2l_1 +1}\;{2 \over 2l_2 +1}\;{2 \over 2l+1} \right]^{1 \over 2}
J_{l_1l_2l} \left [ {2l_1 + 1 \over 2l + 1}\right ]k\; \left [ {2l_2 + 1 \over 2l+1} \right ]k\;
B^{\sg\sgp\sgpp}_{l_1l_2l} \left [ {2l_1 + 1 \over 2l+1}k, {2l_2 + 1 \over 2l + 1}k, k'\right ]
\left ( \begin{array} { c c c }
     l_1 & l_2 & l \\
     s & s' & -(s+s')
  \end{array} \right ).
\ee

\n
Finally if we use the expression for the bispectrum derived using the same approximation, we can write:

\beqa
&& \myC_l^{\sg\sgp,\sgpp}(k,k') =  {[Ac^{-2}]^3} \left [{2 \over 2l+1} \right ] \left [{2l+1 \over 2k} \right ]^2 \left [ {2l+1 \over 2k'} \right ]^2 \sum_{l_1l_2} J^2_{l_1l_2l}
\left ( \begin{array} { c c c }
     l_1 & l_2 & l \\
     0 & 0 & 0
  \end{array} \right )
\left ( \begin{array} { c c c }
     l_1 & l_2 & l \\
     s & s' & -(s+s')
  \end{array} \right )I^{(3)}_{l_1l_2l}\left [{2l+1 \over 2k},{2l+1 \over 2k},{2l+1 \over 2k'}\right ] \nn \\
&& I_{l_1l_2l_3}^{(3)}(r_1,r_2,r_3) = \int_0^{\infty} r^2 dr {\rm B}^{\delta} \left [{l_1 \over r},{l_2 \over r},{l \over r}; r,r,r \right ]
{\cal R}_1[r]{\cal R}_2[r]{\cal R}_3[r]; \quad {\cal R}_i[r]= {F_{\bf K}(r_i,r) \over a(r)}
\label{approx_bicls}
\eeqa

\n
The bispectrum ${\cal B}$ for convergence is defined in Eq.(\ref{eq:con_bi}). The generic symbol
${\cal B}_{l_1l_2l}^{\sg\sgp,\sgpp}$ denotes the bispectrum  for arbitrary spinorial fields as defined in Eq.(\ref{gen_bi}).
Individual cases can be obtained by expressing ${\cal B}_{l_1l_2l}^{\sg\sgp,\sgpp}$ in terms of the convergence
bispectrum using the form factors introduced in Eq.(\ref{individual_bispec}). A numerical implementation of this result may not be
easy due to associated computational cost.

The diagonal elements of the reduced bispectrum $b_{lll}$ are
plotted as a function of $l$ in Fig. \ref{fig:dia_red_bispec}. We
have plotted the projected skew spectrum for convergence
$C_l^{\kappa\kappa,\kappa}$ in Fig. \ref{fig:cl21}. Two different
redshifts were considered $z_s=0.5$ and $z_s=1.0$. The $\Lambda$CDM
background cosmology that we have considered are described by the
following set of parameters: $\Omega_M=0.3$, $\Omega_\Lambda=0.7$,
$\Gamma=0.21$, $h=0.7$ and $\sigma_8=0.90$. We have taken a
hierarchical form for the matter bispectrum and the amplitude $Q_3$
is set to unity. Changing the amplitude $Q_3$ only changes the
overall normalization of the curve. Individual values of
$C_l^{\sg\sgp,\sgpp}$ at a given $l$ depend on the modelling of the
bispectrum for the entire range of $l$ values being considered.
Accurate modelling of the window function is therefore will be an
important ingredient in such calculation. For our study we have
considered a sharp cutoff in the multipole space at $l_{\rm
max}=2000$. The 3D skew spectrum $C_l^{\kappa\kappa,\kappa}$ is
plotted in Fig. \ref{fig:cl21kk} for the same background cosmology.
We plot $C_l^{\kappa\kappa,\kappa}(k,k)$ for three different choice
of $k$ values as function of $l$ (left panel) as well as
$C_l^{\kappa\kappa,\kappa}(k,k)$ for three selection of $l$ values
as a function of the radial wave number $k$.

\subsubsection{The Effect of a Sky Mask}

A mask on the sky is defined through a generic function $w(\oh) =
\sum_{lm}w_{lm}Y_{lm}(\oh)$ on the surface of the sky.  The mask can
be a simple zero and one step function signifying masked and
observed part of the sky or it can also be a more complex apodizing
function with specific weights attach to the different parts of the
sky. We will compute the skew spectrum in the presence of mask and
express it in terms of the skew spectrum in the absence of any mask.
This will allow us to eventually design an estimator which can
estimate the unbiased skew spectrum from the real data in the
presence of a mask and noise. We will consider Gaussian noise, which
means that the estimation of bispectrum is not affected as the noise
has vanishing skew spectrum; however the scatter associated with the
estimator will change in the presence of noise. We will denote the
masked power spectrum by $\tilde C_l$. We will see that convolved or
masked skew spectrum is a linear sum of individual all-sky $C_l$s.
This related to the fact that the use of mask introduces mode-mode
correlation. A matrix inversion with suitable binning can produce
the all-sky $C_l$s from the masked $C_l$s.

In our derivation we start by expanding the masked product field
$[{\gg}({\bf r}){\ggp}({\bf r})w(\oh)]$ in 3D and contract the
resulting harmonics with the masked harmonics of another 3D
spinorial field $[{\ggpp}(\oh)w(\oh)]^*_{lm}$. The 3D harmonic
expansion of the masked product field can be carried out using
harmonics of spin index $s+s'$. Consequently the presence of a
scalar spin-$0$ mask does not alter the spin of the product field
which remains the same as the unmasked product field.

Repeated application of Gaunt's integral allows us to write the
product harmonics in terms of the individual harmonics:

\beqa
 [{\gg}({\bf r}){\ggp}({\bf r})w(\oh)]_{lm}
&=& {\left ( 2 \over \pi \right )}^{3/2}\sum_{l_im_i} J_{l_1l_2l'}J_{l'l_al} (-1)^{l+l'} \sum_{l_am_a} w_{l_am_a} \gg_{l_1m_1}(r)
\ggp_{l_2m_2}(r)  \nn \\
&&  \left (  \begin{array} { c c c }
     l_1 & l_2 & l' \\
     s & s' & -(s+s')
  \end{array} \right ) \left (\begin{array} { c c c }
     l_1 & l_2 & l' \\
     m_1 & m_2 & -m'
  \end{array} \right )
\left (  \begin{array} { c c c }
     l' & l_a & l \\
     (s+s') & 0 &  -(s+s')

  \end{array} \right ) \left (\begin{array} { c c c }
     l' & l_a & l \\
     m' & m_a & -m
  \end{array} \right ).
\eeqa

\n The Fourier-Bessel decomposition of the quadratic field along the
radial direction can be used to relate the 3D harmonics of the
product field with that of individual constituent fields. This will
eventually allow us to express the  3D skew spectrum in terms of the
3D bispectrum. The 3D skew spectrum presented here is a
generalization of our previous work and incorporates full 3D
information. This can be viewed also as a natural generalization of
the 3D power spectrum presented in \cite{Castro05}. Higher-order
counterparts at the level of trispectrum will be presented in the
next section.

\beqa
&& [{\gg}({\bf r}){\ggp}({\bf r})w(\oh)]_{lm}(k)= \nn \\
 \quad\quad && {\left ( 2 \over \pi \right )}^{3/2} \sum_{l_im_i} J_{l_1l_2l'}J_{l'l_al} (-1)^{l+l'} \int dr~ k ~r^2~ j_l(kr)
\sum_{l_am_a} w_{l_am_a} \int dk_1 ~j_l(rk_1)~k_1~ \gg_{l_1m_1}(r)
\int dk_2~j_l(rk_2)~k_2~\ggp_{l_2m_2}(r)  \nn \\
&& \quad \quad \times \left (  \begin{array} { c c c }
     l_1 & l_2 & l' \\
     s & s' & -(s+s')
  \end{array} \right ) \left (\begin{array} { c c c }
     l_1 & l_2 & l' \\
     m_1 & m_2 & -m'
  \end{array} \right )
\left (  \begin{array} { c c c }
     l' & l_a & l \\
     (s+s') & 0 &  -(s+s')
  \end{array} \right ) \left (\begin{array} { c c c }
     l' & l_a & l \\
     m' & m_a & -m
  \end{array} \right ).
\label{square_multipole}
\eeqa

\n
Construction of the masked skew spectrum follows the same approach.
As we pointed out before the 3D skew spectrum
has a radial wave number $k$ dependence built in it. If we integrate the radial dependance
out we can recover the usual projected skew spectrum. Expressing the skew spectrum
in terms of the  multipoles of squared field and individual fields (suitably masked),

\be
\tilde C_l^{\sg\sgp,\sgp}(k,k') = {1 \over 2l+ 1}
\sum_{m=-l}^{l} [{\gg}(\oh){\ggp}(\oh)w(\oh)]_{lm}(k) [{\ggpp}(\oh)w(\oh)]^*_{lm}(k,k').
\label{skew_cl}
\ee

\n
For the 3D harmonic decomposition $[{\ggpp}(\oh)w(\oh)]^*_{lm}(k)$ of individual fields we use the expression
derived previously in Eq.(\ref{individual_multipole}). Using these expressions for the multipoles in the presence of a mask
from Eq.(\ref{square_multipole}) and Eq.(\ref{individual_multipole}) and
in Eq.(\ref{skew_cl}) we can express the skew spectrum constructed  from the masked multipoles or the pseudo skew spectrum
in terms of a coupling matrix and the all-sky skew spectrum in 3D and a mode-coupling matrix. The
mode-mode coupling matrix involves only the mask power spectra $w_l = {1 \over 2l +1}\sum_m w_{lm}w^*_{lm}$, and
depends on the spinorial indices of the respective 3D fields:

\be
G_{ll'}^{ss',s''} = {1 \over 4\pi }\sum_{l_a} {(2l'+1)^2\over (2l+1)}(2l_a+1) \left ( \begin{array}{ c c c }
     l & l_a & l' \\
     s+s' & 0 & -{(s+s')}
  \end{array} \right )
\left ( \begin{array} { c c c }
     l & l_a & l' \\
     s'' & 0 & -s''
  \end{array} \right ) |w_{l_a}|^2; \qquad \qquad s,s',s'' \in0,1, \pm2,3;
\label{skew_mode_coupling_matrix}
\ee

\n
The spin indices $s,s',s''$ take values $0$ for $\kappa$, $\pm2$ for $\gamma_{\pm}$, $-1$ for {\cal F} and
$3$ for {\cal G}. After tedious but straightforward algebra along the line described in \cite{Mu_wl2_10} we can
show that the pseudo-$C_\ell$s expressed as a linear combination of all-sky power spectra can now be expressed
using the following relationship:

\be
\tilde \myC_l^{\sg\sgp,\sgpp}(k,k') = \sum_{l'} G_{ll'}^{ss',s''}
\myC_{l'}^{\sg\sgp,\sgpp}\left({2l+1 \over 2l'+1}k,{2l+1 \over 2l'+1}k' \right).
\ee
The deconvolution process to recover the all-sky $\myC_{l'}^{\sg\sgp,\sgpp}(r_1,r_2)$ from its masked
counterpart involves inversion of the coupling matrix $G$ which depends not only $l$ indices but
also on spinorial indices associated with the fields. In case of small-sky coverage which might
be the case for present generation of surveys, for inversion of the matrix will typically involve
binning to avoid any possible numerically-singular matrices.

Here it is worth pointing out that the various spinorial fields that we can use to construct
individual skew spectrum do probe the same underlying matter bispectrum. This can be used as a helpful
diagnositic to probe possible spurious effects of mask and noise.

The power spectrum $\myC_l^{\sg \sgp, \sgpp}(k,k')$ reported here is
an extension of similar power spectrum introduced in \cite{MuHe09}
for CMB studies. Later it was used in \cite{MuHeCo_wl1_10} to probe
the convergence skew spectrum and was also generalized for projected
spin-skew spectrum in \citep{Mu_wl2_10}. In this study we present
skew spectrum for spinorial fields using a complete 3D analysis. The
power spectrum is specified by multipole indices on the surface of
the celestial sphere as well as radial harmonics along the line of
sight direction. Results are generic for fields with arbitrary spin
and can be used to probe shear, flexion or convergence maps. Similar
statistics in the coordinate space has been reported before.
\cite{berludoMell03} studied $\langle \gamma^2(\oh)\gamma(\oh')
\rangle$ which directly deals with shear maps as opposed to
convergence maps. This statistics along with a similar but simpler
version which uses $\langle\kappa^2(\oh)\kappa(\oh')\rangle$ was
also studied. The perturbation theory was employed to model the
underlying mass distribution and it used a flat sky approximation to
simplify their calculations. A complementary statistic $\langle
[\gamma(\oh_1)\cdot\gamma(\oh_2)]\gamma \rangle$ was also considered
which relies on more detailed modelling of the bispectrum. These
statistics were used by \cite{BerVanMell02} later to detect
non--Gaussianity from the VIRMOS-DESCART Lensing Survey.

Our results presented here deal with power spectrum associated with
the higher-order multispectra, are derived using generic all-sky
treatment, and can also handle decomposition into Electric and
Magnetic components in a much more straightforward manner.
The results presented here are not only applicable
to shear or convergence but are also applicable for higher-order spinorials such as {\it Flexions}.
To increase the signal-to-noise of the estimates it is customary to often sum all possible
mode of $C_l^{\sg\sgp,\sgpp}(k,k)$ in to a single number which is called {\it skewness}.

\be
S_3^{\sg\sgp,\sgpp}(k,k') = \sum_l (2l+1) \myC_l^{\sg\sgp,\sgpp}(k,k')
\label{skewness}
\ee

\n
For a concrete expression for $S_3^{\sg\sgp,\sgpp}(k,k')$ we need to replace $\myC_l^{\sg\sgp,\sgpp}(k,k')$
by the expression derived in Eq.(\ref{approx_bicls}) or Eq.(\ref{exact_bicls}).

To make connection with the real space statistics we can use the following relation:

\be
\myC_l^{\sg\sgp,\sgpp}(k,k') = { 2 \over \pi } \int dk k r^2 j_l(kr)  \int dk' k' r'^2 j_l(k'r')~\myD_l^{\sg\sgp,\sgpp}(r,r'),
\label{clcl}
\ee

\n
and a similar relation can be derived for the skewness $S_3^{\sg\sgp,\sgpp}(r,r')$ and $S_3^{\sg\sgp,\sgpp}(k,k')$
defined above. One point statistics such as $S_3(k,k')$ only contain radial information as all spherical harmonics
are already summed over.

\n If we make the approximation of replacing the spherical Bessel
function with a delta-function form we can write \be
\myC_l^{\sg\sgp,\sgpp}(k,k') = {2 \over 2l+1}
\myD_l^{\sg\sgp,\sgpp}\left [ {2l+1 \over 2k},{2l+1 \over 2k'}
\right ]. \quad
\label{approx_clcl} \ee An equivalent expression is valid for
$S_3^{\sg\sgp,\sgpp}(k,k')$. In both Eq.(\ref{clcl}) and
Eq.(\ref{approx_clcl}) we use the same notations to define the
harmonic space $\myC_l^{\sg\sgp,\sgpp}(k,k')$ and real space
$\myC_l^{\sg\sgp,\sgpp}(r,r')$ power spectra. However their
functional dependence on their arguments are different. It is
interesting to notice that in Fourier domain the
$\myC_l^{\sg\sgp,\sgpp}(k,k')$ probes specific length scales.

\subsection{The Kurt spectrum}

The four-point correlation function, or its harmonic counterpart the
trispectrum, has been extensively studied in the literature . This
contains the information about the non--Gaussianity beyond the
lowest level \citep{Hu99,OkaHu02}.

For the case of weak lensing studies clearly the gravity-induced
non--Gaussianity is the main motivation. Studies in trispectrum
analysis have also been pursued using various other probes e.g.
using 21-cm surveys \citep{CooLiMel08} or more extensively in
several CMB studies; see \citet{BART04} for a review. However these
studies typically probe the trispectrum induced by primordial
non--Gaussianity. It is important to note that at the level of
four-point, the Gaussian part of the signal as well as the noise
both carry a non-zero (unconnected) trispectrum. This degrades the
signal-to-noise for various estimators and clearly needs to be
subtracted out before an unbiased comparison with the theoretical
predictions can be made. It is obvious that the detection of the
trispectrum from noisy data is far more nontrivial than the
estimation of the bispectrum.
We provide
analytical expressions here mainly for completeness and to show that the generic results can be
obtained based on very simple hierarchical ansatz.

Previous studies have mainly concentrated on one-point estimators
which collapse the data to a single number - known as the kurtosis.
We extend studies involving kurtosis
$\langle[\gg\ggp\ggpp\ggppp](\oh,r)\rangle$ to its two-point
counterparts: $\langle[\gg\ggp](\oh,r)[\ggpp\ggppp](\oh',r')\rangle$
and $\langle \sg\sgp\sgpp(\oh,r)\sgppp(\oh',r')\rangle$. In practice
however we will consider the Fourier transforms of these objects in
3D, \cg{the kurt spectra}, which are the power spectra associated
with the trispectra, $\myC_l^{(\sg\sgp,\sgpp\sgpp)}(k,k')$ and
$\myC_l^{(\sg\sgp,\sgpp\sgppp)}(k,k')$. As was the case for the skew
spectrum, we not only do harmonic decomposition on the surface of
the celestial sphere but on as well on the radial direction. For the
construction of  $\myC_l^{(\sg\sgp,\sgpp\sgppp)}(k,k')$.

\subsubsection{Two-to-Two Kurt Spectrum}

The first of two kurt spectra $\myC_l^{(\sg\sgp,\sgpp\sgppp)}(k,k')$
can be constructed from the harmonic transform of
$[\gg\ggp]_{lm}(k)$ of the quadratic combination of two arbitrary
spin-weight fields discussed previously in the context of the skew
spectrum Eq.(\ref{eq:decompose}). The resulting kurt spectrum is
generic and can be defined for any given set of four spin-weight
functions defined in 3D. Unlike the skew spectrum which is zero for
Gaussian fields, the kurt spectra are non-zero even in the absence
of any non--Gaussianity, which introduces additional complexity. The
Gaussian contribution (also known as the disconnected piece) needs
to be subtracted out before it is employed for the study of
non--Gaussianity. The noise, often assumed Gaussian, can also be
subtracted following the same technique. It will contribute only to
the  disconnected part. Later in this section, we will also consider
the effect of a mask as we did for the skew spectrum.

We will use these results to derive expressions for $\myC_l^{(\sg\sgp\sgpp,\sgppp)}(k,k')$ which leads to:
$\myC_l^{(\sg\sgp,\sgpp\sgpp)}(k,k') = {1 \over 2l+1} \sum_m [\gg\ggp]_{lm}^{*}(k) [\ggpp\ggpp]_{lm}(k').$
These power spectra directly probe $T_{l_3l_4}^{l_1l_2}(l,k,k')$. It compresses all the
available information in quadruplets of modes specified by
$(l_1,l_2,l_3,l_4)$ to a power spectrum. The power spectra
$\cred{\myC}_l^{(\sg\sgp,\sgpp\sgppp)}(k,k')$ and $\cred{\myC}_l^{(\sg\sgp\sgpp,\sgppp)}(k,k')$ differ in the way
they associate weights to various modes and contain complementary information.
The reduced trispectrum $T_{l_1l_2}^{l_3l_4}(k_i;L)$ is defined in terms of
$\langle \gg_{l_1m_1}(k_1) \ggp_{l_2m_2}(k_2) \ggpp_{l_3m_3}(k_3) \ggppp_{l_4m_4}(k_4) \rangle_c$ as follows.
We have added the radial distances $r_i$ associated with each spherical harmonic
in the argument with $L$, which specifies the diagonal formed by the quadruplet of
four quantum numbers $l_i$ $\langle \gg_{l_1m_1}(k_1) \ggp_{l_2m_2}(k_2)
\ggpp_{l_3m_3}(k_3)\ggppp_{l_4m_4}(k_4) \rangle_c = \sum_{LM} (-1)^M
T^{l_1l_2}_{l_3l_4}(k_i;L) \left ( \begin{array}{ c c c
}
     l_1 & l_2 & L \\
     m_1 & m_2 & M
  \end{array} \right)
\left ( \begin{array}{ c c c }
     l_3 & l_4 & L \\
     m_3 & m_4 & -M
  \end{array} \right).$ The final expression depends on the spin indices of various fields as well
as on a Kernel $F$ (defined below) which has angular harmonic numbers $l_i$ and radial wave numbers $k,k'$
as its arguments:

\be \myC_l^{(\sg\sgp,\sgpp\sgppp)}(k,k') =
\frac{1}{(2l+1)^2}\sum_{l_1l_2l_3l_4}  {\cred S}_{l\cred l_1l_2}{\cred S}_{ll_3l_4}
\left ( \begin{array}{ c c c }
     l_1 & l_2 & l \\
     s_1 & s_2 & -(s_1+s_2)
  \end{array} \right)
\left ( \begin{array}{ c c c}
     l_3 & l_4 & l \\
     s_3 & s_4 & -(s_3+s_4)
  \end{array} \right)
F^{(2,2)}(l_i,l,k,k'). \label{sphere22} \ee

\n
The kernel $F^{(2,2)}(l_i,l,k,k')$ is defined in terms of the reduced trispectrum $T_{l_1l_2}^{l_3l_4}(k_i;L)$.

\beqa
F^{(2,2)}(l_i,l,k,k') = && \left ({2 \over \pi} \right )^{3} \int_0^{\infty}dr~r^2~k~j_l(kr)\int_0^{\infty}dr'~r'^2~k'~j_l(k'r') \nn \\
&& \times \int_0^{\infty} dk_1 k_1j_l(k_1r)\int_0^{\infty} dk_2 k_2j_l(k_2r)
\int_0^{\infty} dk_3 k_3j_l(k_3r')\int_0^{\infty} dk_4 k_4j_l(k_4r')
T^{l_1l_2}_{l_3l_4}(k_i;L).
\eeqa

\begin{table*}
\caption{Notations}
\begin{center}
\begin{tabular}{|c |c| c}
\hline
\hline
Power Spectrum (R/F) & $\myD_l(r_1,r_2), \myC_l(k_1,k_2) $ & Eq.(\ref{eq:cl21})  \\
Variance (R/F) & $\Sigma(r_1,r_2), \Sigma(k_1,k_2)$ & Eq.(\ref{var}) \\
\hline
\hline
Bispectrum & $B^{\delta}_{l_1l_2l_3}$, ${\cal B}^{\kappa\kappa\kappa}_{l_1l_2l_3}$,~${\cal B}^{\sg\sgp\sgp}_{l_1l_2l_3}$ & Eq.(\ref{bi_def}),~Eq.(\ref{individual_bispec})\\
\hline
Skew Spectrum (R/F) & $\myD_l^{\sg\sgp,\sgpp}(r_1,r_2), \myC_l^{\sg\sgp,\sgpp}(k_1,k_2)$  & Eq.(\ref{exact_bicls})
Eq.(\ref{approx_bicls}) \\
Skewness (R/F) & $S_3^{\sg\sgp\sgpp}(r_1,r_2), S_3^{\sg\sgp\sgpp}(k_1,k_2)$ & Eq.(\ref{skewness}), Eq.(\ref{clcl}) \\
\hline
\hline
Trispectrum & ${}^{\delta}T^{l_1l_2}_{l_3l_4}$, ${}^{\kappa}{\cal T}^{l_1l_2}_{l_3l_4}$ & Eq.(\ref{eq:tri}) \\
\hline
Kurt Spectrum (R/F) & $\myD_l^{\sg\sgp,\sgpp\sgppp}(r_1,r_2), \myC_l^{\sg\sgp,\sgpp\sgppp}(k_1,k_2)$  & Eq.(\ref{2to1}) \\
 & $\myD_l^{\sg\sgp\sgpp,\sgppp}(r_1,r_2), \myC_l^{\sg\sgp\sgpp,\sgppp}(k_1,k_2)$  & Eq.(\ref{3to1}) \\
Kurtosis(R/F) & $K_4^{\sg\sgp,\sgpp\sgppp}(r_1,r_2),  K_4^{\sg\sgp\sgpp,\sgppp}(k_1,k_2)$ & Eq.(\ref{k4}) \\
\hline
\hline
Deconvolution, Mixing Matrix & $\tilde C_l$, $\hat C_l$, $M_{ll'}, G_{ll'}$  & Eq.(\ref{mode_coupling_cls}),
~ Eq.(\ref{ps_EB}),~ Eq.(\ref{skew_mode_coupling_matrix}),~ Eq.(\ref{mm})  \\
\hline
\hline
\end{tabular}
\end{center}
\label{table:notation}
\end{table*}

\n
We will consider the two components of the trispectrum ``snake'' and ``stars'' separately for each
of the two kurt specrta. If we follow the
algebra, which is very similar to what was done previously to derive the skew spectrum we arrive at
the following expression for the star component of the three-to-one kurt spectrum. The expression reduces
to an one dimensional integral as we use the Limber approximation for simplification.

\ben
\myC_l^{(\sg\sgp,\sgpp\sgppp)}(k,k') &=& {A^4 \over c^8}{\left [ 2 \over 2l+1 \right ]} {\left [ 2l+1 \over 2k \right]^2}
{\left [ 2l+1 \over 2k' \right ]^2}
\frac{1}{(2l+1)^2}\nn \\
&& \times \sum_{l_1l_2l_3l_4}{\cred S}_{l\cred l_1l_2}{\cred S}_{ll_3l_4}
\left ( \begin{array}{ c c c }
     l_1 & l_2 & l \\
     s_1 & s_2 & -(s_1+s_2)
  \end{array} \right)
\left ( \begin{array}{ c c c}
     l_3 & l_4 & l \\
     s_3 & s_4 & -(s_3+s_4)
  \end{array} \right)
I^{(4)}_{tar} \left [{2l + 1 \over 2k },{2l+1 \over 2k},{2l+1 \over 2k'},{2l+1 \over 2k'} \right ]
\label{2to1} \een The contribution from the {\it star} diagram can
be expressed in a similar manner. We simply have to replace the
$I^{(4)}_{\rm snake}$, which is defined in Eq.(\ref{snake}) with
$I^{(4)}_{\rm star}$ defined in Eq.(\ref{star}). The additional
terms that describe the exchange symmetry of snake terms however
will involve the computation of $6j$ symbols Eq.(\ref{6j}) which
poses additional computational complexity. There are two cumulant
correlators at four-point level as explained above.

\subsubsection{Three-to-One kurt spectrum}

 Following the
discussion above we now focus on the  other degenerate power spectra associated with the
cumulant correlator $\langle[\gg\ggp\ggpp](\oh)\ggppp(\oh')\rangle$.We will start by defining the all-sky harmonic
transform  $[\sg\sgp\sgpp]_{lm}(k)$ for the cubic field $\gg\ggp\ggpp(\oh,r)$ and cross-correlate it against the
remaining field $\ggppp_{lm}(k')$.  This is of the same order
as $\langle[\gg\ggp](\oh)[\ggpp\ggppp](\oh') \rangle$ and contains information about
trispectra as well. The compression of the information is done with different
weighting for different modes: $\myC_l^{(\sg\sgp\sgpp,\sgppp)}(k,k') = {1 \over 2l+1} \sum_m \mathrm {Real}
\left \{ [\ggp\ggpp\ggppp]_{lm}^{*}(k) \ggp_{lm}(k') \right \}$. We can now use the definition of
the trispectra $T^{l_1l_2}_{l_3l}(L;k_1,k_2)$ to express
$\cred{\myC}_l^{(\sg\sgp\sgpp,\sgppp)}(k_1,k_2)$ in terms of the trispectra. The main difference with the previous
spectrum $\cred{\myC}_l^{(\sg\sgp,\sgpp\sgppp)}(k_1,k_2)$ is that it sums over all possible configurations
of the quadrilateral keeping one of the sides fixed, whereas $\myC_l^{(\sg\sgp,\sgpp\sgppp)}(k_1,k_2)$
keeps one of the diagonals fixed but \cg {otherwise} sums over all possible configuration of the
quadrilateral.  The harmonics of the cubic field can be expressed in terms of the individual harmonics using the
following expressions. In the first equation we treat the radial direction using real-space expression
and next we also do a Fourier transform in the radial direction to obtain a full 3D decomposition.

\ben
&& [\gg\ggp\ggpp]_{lm}(r) = \sum_{l_im_i}\gg_{l_1m_1}(r)\ggp_{l_2m_2}(r)\ggpp_{l_3m_3}(r)
\int d \hat \Omega Y_{l_1m_1}(\hat \Omega)Y_{l_2m_2}(\hat \Omega)Y_{l_3m_3}(\hat \Omega)Y^*_{lm}(\hat \Omega) \nn \\
&&  [\gg\ggp\ggpp]_{lm}(k)= \left ({2 \over \pi} \right )^{2} \int_0^{\infty}dr~r^2~k~j_l(kr)
\sum_{l_im_i} \int_0^{\infty} dk_1 k_1j_l(k_1r)\int_0^{\infty} dk_2 k_2j_l(k_2r)
\int_0^{\infty} dk_3 k_3j_l(k_3r) \gg_{l_1m_1}(k_1)\ggp_{l_2m_2}(k_2)\ggpp_{l_3m_3}(k_3). \nn \\
\een

\n
Following these expression we can express the three-to-one kurt spectra in terms of the trispectrum.
The geometric prefactors that appear in the projected (2D) three-to-one power spectra also appear
in the 3D expression. The radial harmonics dependence comes through the kernel $F^{(3,1)}$:

\be \myC_l^{(\sg\sgp\sgpp,\sgppp)}(k_1,k_2) =
\frac{1}{2l+1}\sum_{l_1l_2l_3;L}\frac{1}{2L+1}{\cred S}_{l_1l_2L}{\cred S}_{Ll_3l}
\left ( \begin{array}{ c c c }
     l_1 & l_2 & L \\
     s_1 & s_2 & -(s_1+s_2)
  \end{array} \right)
\left ( \begin{array}{ c c c}
     L & l_3 & l \\
     (s_1+s_2) & s_3 & -(s_1+s_2+s_3)
  \end{array} \right)F^{(3,1)}(l_i,l,k,k')
\label{sphere31} \ee

\n
The kernel $F^{(3,1)}(l_i,l,k,k')$ is defined in terms of the trispectrum $T_{l_1l_2}^{l_3l_4}(k_i;L)$
as follows:

\be
F^{(3,1)}(l_i,l,k,k') = \left ({2 \over \pi} \right )^{3} \int_0^{\infty}dr~r^2~k~j_l(kr)
\int_0^{\infty} dk_1 k_1j_l(k_1r)\int_0^{\infty} dk_2 k_2j_l(k_2r)\int_0^{\infty} dk_3 k_3j_l(k_3r')
T^{l_1l_2}_{l_3l_4}(k_i;L).
\label{def_F31}
\ee

\n
The star contribution can be expressed in terms of the kernel
$I^{(4)}_{\rm star}$, using the Limber approximation: \ben &&
\myC_l^{(\sg\sgp\sgpp,\sgppp)}(k,k') = {A^4 \over c^8}{\left [ 2
\over 2l+1 \right ]} {\left [ 2l+1 \over 2k \right]^2} {\left [ 2l+1
\over 2k' \right ]^2}
\frac{1}{(2l+1)^2}\nn \\
&& \times \sum_{l_1l_2l_3l_4}{\cred S}_{l\cred l_1l_2}{\cred S}_{ll_3l_4}
\left ( \begin{array}{ c c c }
     l_1 & l_2 & L \\
     s_1 & s_2 & -(s_1+s_2)
  \end{array} \right)
\left ( \begin{array}{ c c c}
     L & l_3 & l \\
     s_1 + s_2 & s_3 & -(s_1+s_2+s_3)
  \end{array} \right)
I^{(4)}_{\rm star} \left [{2l + 1 \over 2k },{2l+1 \over 2k},{2l+1
\over 2k},{2l+1 \over 2k'} \right ].
\label{3to1}
\een

\n For the {\em snake} component, we need to replace the kernel
$I^{(4)}_{\rm star}$ with $I^{(4)}_{\rm snake}$ which also involves
a single integration. The evaluation of {\em snake} terms are
however difficult as in addition there will be terms that involve 6j
symbols Eq.(\ref{6j}) because of their permutation symmetry. As was
the case with skew spectrum, we can collapse these fourth order
two-point objects to reduce them to one-point numbers, the kurtosis,
which will be a function of both radial harmonics $k_1,k_2$:

\be
K_4^{\sg\sgp,\sgpp\sgppp}(k_1,k_2) = \sum_l (2l+1) \myC_l^{(\sg\sgp,\sgpp\sgpp)}(k_1,k_2); \quad
K_4^{\sg\sgp\sgpp,\sgppp}(k_1,k_2) = \sum_l (2l+1) \myC_l^{(\sg\sgp\sgpp,\sgppp)}(k_1,k_2).
\label{k4}
\ee

\n
The corresponding real space (in radial direction)  versions  $K_4^{\sg\sgp,\sgpp\sgppp}(r_1,r_2)$,
and $K_4^{\sg\sgp\sgpp,\sgppp}(r_1,r_2)$  can be related using an equation similar to
Eq.(\ref{clcl}).

We will deal with the mode coupling issues arising from the partial
sky coverage next. We will show how to deconvolve the effect of
mask. It is important to keep in mind that the one point objects
such as $K_4^{\sg\sgp,\sgpp\sgppp}(k_1,k_2)$ will have more
signal-to-noise and can play important role in studying growth of
non--Gaussianity along the radial direction.

A few general comments are in order. At the level of the trispectrum there are two hierarchical amplitudes.
If we employ the two sets of kurt spectra then the amplitudes can be determined. This is very similar to
their use in the CMB where the kurt spectra can be used to provide independent constraints
for parameters $g_{NL}$ and $\tau_{NL}$ that describe the Taylor expansion of the
inflationary potential \citep{Hu99,OkaHu02}. Indeed in some hierarchical models the {\em snake}
amplitude $R_a$ and the bispectrum amplitude are related by $R_a = Q^2$. Similar relation
also exists between $f_{NL}$ and $\tau_{NL}$. The study of skew spectrum can provide
direct estimates of the parameter $Q_3$. The estimates from skew spectrum is
expected to be more significant statistically due to the higher signal-to-noise.
An independent estimation using kurt spectra can provide a direct test of various
hierarchical ansatz. It is important to realize that the skew spectra as well as
the kurt spectra are integrated quantities, i.e. their amplitudes at a specific $l$ depends
on the entire range of $l$ values being considered. In terms of modelling of these quantities,
it means that any successful theoretical prediction will have to correctly model the relevant multi-spectra
for the entire range of $l$ values being probed. The procedure can be extended to even higher-order.
The number of distinct topological diagrams that are needed to build a correlation function
at a given order will be same as the number of related power spectra, e.g. at fifth order
there are three topological amplitudes and three multispectra. The procedure
outlined above can be extended in such cases as signal to noise of the available data improves.

\subsubsection{The effect of a Mask and Subtraction of Gaussian Contribution}

The partial sky coverage will mean that the measured power spectrum
$\tilde \myC_l^{(\sg\sgp\sgpp,\sgppp)}(k_1,k_2)$ is not the same as
theoretical expectation, but is related as before by: $\tilde
\myC_l^{(\sg\sgp\sgpp,\sgppp)}(k,k')=G_{ll'}\myC_{l'}^{(\sg\sgp\sgpp,\sgppp)}(k,k')$.
In fact it can shown that for arbitrary sky coverage with arbitrary
mask the above analysis can be generalized to arbitrary order of
correlation hierarchy. If we consider a correlation function at
$p+q$ order, for every possible combination of $(p,q)$ we will have
an associated power spectrum. Using the same expression for the mode
mixing matrix, we can invert the observed ${\tilde
\myC}_{l}^{(p,q)}(k_1,k_2)$ to ${\hat \myC}_{l}^{(p,q)}(k_1,k_2)$.
Hence for an arbitrary mask with arbitrary weighting functions the
deconvolved set of estimators can be written as: ${\hat
\myC}_l^{(p,q)}(k_1,k_2) = \sum_{l'}G_{ll'}^{-1} {\tilde
\myC}_{l'}^{(p,q)} \left ({2l'+1 \over 2l+1}k_1,{2l'+1 \over
2l+1}k_2 \right )$. The matrices $G$ (defined below) depend on the
spin indices as well as the power spectrum of the mask.

\beqa
&& G_{ll'}^{ss',s''s'''} = {1 \over 4\pi }\sum_{l_a} {(2l'+1)^2\over 2l+1} (2l_a+1) \left ( \begin{array}{ c c c }
     l & l_a & l' \\
     s+s' & 0 & -{(s+s')}
  \end{array} \right )
\left ( \begin{array} { c c c }
     l & l_a & l' \\
     s''+s''' & 0 & -(s''+s''')
  \end{array} \right ) |w_{l_a}|^2; \nn \\
&& G_{ll'}^{ss's'',s'''} = {1 \over 4\pi }\sum_{l_a} {(2l'+1)^2 \over 2l+1}(2l_a+1) \left ( \begin{array}{ c c c }
     l & l_a & l' \\
     (s+s'+s'') & 0 & -{(s+s'+s'')}
  \end{array} \right )
\left ( \begin{array} { c c c }
     l & l_a & l' \\
     s''' & 0 & -s'''
  \end{array} \right ) |w_{l_a}|^2; \qquad s,s',s'',s''' \in0,1, \pm2,3.\nn \\
\label{mm}
\eeqa

However, as pointed out before, if we define kurt spectra
${\hat \myC}_l'^{(p,q)}(k_1,k_2)= (2l+1){\hat \myC}_l^{(p,q)}(k_1,k_2)$  we can use
the same mode coupling matrices that are used in projection for the purpose of
deconvolution. However the 3D treatment introduces a remapping of the radial
mode due to the presence of a mask ${\tilde \myC}_l'^{(p,q)}(k_1,k_2)=
\sum_{l'} M_{ll'}{\myC}_l'^{(p,q)}\left ({2l'+1 \over 2l+1 }k_1,{2l'+1 \over 2l +1}k_2 \right )$.

The Gaussian contribution to the trispectrum ${\cal G}$ can be written as:

\ben
{\cal G}^{l_1l_2}_{l_3l_4}(k_1,k_2,k_3,k_4;L) = && (-1)^{l_1+l_3} \sqrt {(2l_1+1)(2l_3+1)} \myC^{\sg,\sgp}_{l_1}(k_1,k_2)\myC_{l_3}^{\sgpp,\sgppp}(k_3,k_4) \delta_{L0}\delta_{l_1l_2}\delta_{l_3l_4} \nonumber \\
&&  + (2L+1) (-1)^{l_2+l_3+L} \delta_{l_1l_3}\delta_{l_2l_4} \myC_{l_1}^{\sgp\sgpp}(k_1,k_3)
\myC_{l_2}^{\sgpp\sgppp}(k_2,k_4)
  +  (2L+1)\myC_{l_1}^{\sg,\sgppp}(k_1,k_4)\myC_{l_2}^{\sgp,\sgpp}(k_2,k_3)  \delta_{l_1l_4}\delta_{l_2l_3}.
\een

\n The various power spectra $\myC_{l_1}^{\sgp\sgpp}(k_1,k_3)$ above
include contributions from signal and noise. Next we can compute the
Gaussian contributions to $\myC_l^{(\sg\sgp\sgpp,\sgppp)}$  and
$\myC_l^{(\sg\sgp,\sgpp\sgppp)}$ following the same procedure as
before just by replacing the trispectrum $T^{l_1l_2}_{l_3l_4}$ with
its Gaussian counterpart $G^{l_1l_2}_{l_3l_4}(k_i;L)$. Indeed we
will have to keep the ordering correct for various $l_i$ and their
$r_i$ counterparts. It is also important to realize that in
computing the Gaussian contribution we will have to take into
account both the signal and the noise $\myC_ls$ (assumed to be
Gaussian), i.e, $\myC_l= \myC_l^S+\myC_l^N$. We will denote the
Gaussian contributions to the (three-to-one) kurt-spectra by ${\cal
G}_l^{(\sg\sgp\sgpp,\sgppp)}(k_1,k_2)$ and (two-to-two) ${\cal
G}_l^{(\sg\sgp,\sgpp\sgppp)}(k_1,k_2)$ respectively:

\beqa
&& {\cal G}_l^{(\sg\sgp\sgpp,\sgppp)}(k_1,k_2)
= \frac{1}{2l+1}\sum_{l_1l_2l_3;L}\frac{1}{2L+1}{\cred S}_{l_1l_2L}{\cred S}_{Ll_3l} {\cal G}^{l_1l_2}_{l_3l}(L,k_1,k_2); \nn \\
&& {\cal G}_l^{(\sg\sgp,\sgpp\sgppp)}(k_1,k_2) = \frac{1}{(2l+1)^2}\sum_{l_1l_2l_3l_4}
{\cred S}_{ll_1l_2}{\cred S}_{ll_3l_4} {\cal G}_{l_3l_4}^{l_1l_2}(l,k_1,k_2) .
\eeqa

\n For realistic surveys with a mask, the unconnected (Gaussian)
contributions to the total kurt spectrum, listed above, can be
deconvolved in a manner identical to what we have presented before
for the connected part of the total trispectrum. The mode mixing
matrix for the Gaussian contribution is identical to what we have
introduced in Eq.(\ref{mm}). From the estimated $\tilde
\myD_l^{(\sg\sgp\sgpp,\sgppp)}(r_1,r_2)$ and $\tilde
\myD_l^{(\sg\sgp,\sgpp\sgppp)}(r_1,r_2)$  these contributions need
to be subtracted out before comparing them against the theoretical
expectations $\myC_l^{\sg\sgp,\sgpp\sgppp}(k,k') = {2 \over 2l+1}
\myD_l^{\sg\sgp,\sgpp\sgppp}\left [ {2l+1 \over 2k},{2l'+1 \over
2k'} \right ]$. An equivalent expression holds for the Gaussian
contributions that needs to be subtracted: ${\cal
G}_l^{\sg\sgp,\sgpp\sgpp}(k_1,k_2)$ and ${\cal
G}_l^{\sg\sgp,\sgpp\sgpp}(k_1,k_2)$.

We have so far only considered the gravity-induced trispectrum in
our discussion. However the kurt spectra for primordial
non--Gaussianity can be derived in a very similar manner by
replacing the gravity-induced trispectrum with a corresponding model
for the primordial trispectrum. However it is expected that
gravity-induced non--Gaussianity will dominate the primordial ones
at least at lower redshift.

\section{Error Analysis}

In the previous section, we have derived the expression for the 3D power spectrum associated with convergence
field and indicated how a similar analysis can be performed for other spinorial fields.
Estimation of these power spectra from noisy data will however will always have to deal
with issues such as noise and partial sky coverage. An estimator which can deal with
such observational constraints was developed for the case of power spectra in Eq.(\ref{mode_coupling_cls}).
It is however clear that the estimation of power spectra to be meaningful for
any cosmological study we need an approximate handle on the error-bars and their
covariance.

The error analysis for the PCL estimator that was introduced, can be
done using the formalism used in \cite{Mucross} which is based on pseudo-Cls
formalism developed by various authors in the context of CMB data analysis \citep{Efs1,Efs2,BCT}. The contributions to the error
covariance will have three different
components. On large angular scales or small $l$ the error will be dominated
mainly by cosmic variance where as the high $l$ or smaller angular scale it will
mainly be dominated by noise due to the intrinsic ellipticity of galaxies.

\subsection{Power spectrum}

\n
$\myC_l^{\sg,\sgp}(k,k')$ defines the cross-spectra between two spinorial fields $\gg(\oh)$ and $\ggp(\oh)$.  That is
$\tilde \myC_l^{\sg,\sgp}(k,k') = {1 \over 2l+1}\sum_m {\rm Re}[_s{\tilde {\sg}^{lm*}(k)
_{s'}\tilde\sgp_{lm}(k')}].$ It is possible to derive the covariance of estimates under certain simplifying
assumptions. The general principles for deriving these results are outlined
in \cite{Mucross} and will not be repeated here, and we quote the results for
the ordinary power spectra here. For simplicity we will only consider $s=s'$. In later sub-sections we will also
consider covariance matrices for the skew spectrum. The covariance matrix of the estimates is
$\langle \delta \myC_l^{\sg,\sgp}(k)\delta \myC_{l'}^{\sg,\sgp}(k')\rangle$, where $\delta \myC_l^{\sg,\sgp}(k)$
are the deviations from the ensemble average $\langle  \myC_l^{\sg,\sgp}(k) \rangle$:

\beqa
&& \langle \delta \myC_l^{\sg,\sgp}(k,k')\delta \myC_{l'}^{\sg,\sgp}(k,k')\rangle \approx \Sigma^{SS}_{ll'}(k,k') +
\Sigma^{SN}_{ll'}(k,k') + \Sigma^{NN}_{ll'}(k,k') \nn\\
&&\Sigma^{SS}_{ll'}(k,k') = {1 \over 4\pi}\left \{ {\sqrt{\myC_l^{\sg,\sg}(k,k)\myC_{l'}^{\sg,\sg}(k,k)}
\sqrt{\myC_l^{\sgp,\sgp}(k',k')\myC_{l'}^{\sgp,\sgp}(k',k')} +
\myC_l^{\sg,\sgp}(k,k')\myC_{l'}^{\sg,\sgp}}(k,k') \right \}
 \sum_{l_{\alpha}m_{\alpha}}
\left ( \begin{array} { c c c }
     l & l_a & l' \\
     s & 0 &  -s'
  \end{array} \right )^2 |w_{l_a}|^2 \nn \\
&&\Sigma^{NN}_{ll'}(k,k') =
{1 \over 4 \pi}\sum_{l_{\alpha}m_{\alpha}}
\left ( \begin{array} { c c c }
     l & l_a & l' \\
     s & 0 &  -s'
  \end{array} \right )^2
\left \{ |[\sigma^{\sg,\sg}w]_{l_{\alpha}m_{\alpha}}(k,k)
[\sigma^{\sgp,\sgp}w]^*_{l_{\alpha}m_{\alpha}}(k',k')|
+ |[\sigma^{\sg,\sgp}w^2]_{l_{\alpha}m_{\alpha}}(k,k')|^2 \right \}\nn \\
&&\Sigma^{SN}_{ll'}(k,k') =  {1 \over 4\pi}\sum_{l_{\alpha}m_{\alpha}}
\left ( \begin{array} { c c c }
     l & l_a & l' \\
     s & 0 &  -s'
  \end{array} \right )^2
\Big \{ |[\sigma^{\sgp,\sgp}w]_{l_{\alpha}m_{\alpha}}(k',k')[w]_{l_{\alpha}m_{\alpha}}|
\sqrt{\myC_l^{\sg,\sg}(k,k)\myC_{l'}^{\sg,\sg}(k,k)} + {\rm symm.\ term.} \nn \\
&& \quad\quad\quad\quad\quad\quad\quad\quad\quad\quad\quad\quad + 2\;|[\sigma^{\sg,\sgp}w]_{l_{\alpha}m_{\alpha}}(k,k')
[w]_{l_{\alpha}m_{\alpha}}|
\sqrt{\myC_l^{\sg,\sgp}(k,k')\myC_{l'}^{\sg,\sgp}(k,k')} \Big \}.
\eeqa

\n
The symm. term. can be constructed by exchanging $k$ and $k'$
as well as $\sg$ and $\sgp$. We have divided the total contribution into three different
components. The term $\Sigma^{SS}_{ll'}(k,k')$ is the cosmic variance contribution and
depends on the target spectra but is independent of noise. The term $\Sigma^{NN}_{ll'}(k,k')$
signifies the noise contribution and finally $\Sigma^{SN}_{ll'}(k,k')$ is a cross term
which gets contributions from both signal and noise.
We have assumed that the noise is statistically uncorrelated but it varies with pixel-position in the sky; i.e.
 $\langle\sg(\oh,r)\sgp(\oh',r')\rangle=\sigma^{\sg,\sgp}(\oh,r,r')\delta_{2D}(\oh-\oh')$.

A detailed modelling of source distribution is required for 3D error
estimates. The observational mask $w(\Omega)$ that we use is
completely generic however our results uses completeness and
orthogonality of spherical harmonics on the cut-sky. This means
results will be accurate only for near all-sky coverage. The various
window functions that we have introduced are constructed from the 3D
harmonic transforms such as $[\sigma^2w^2]_{LM}(k)$ and $[\sigma
w]_{LM}(k)$ of maps constructed from {\em 3D noise maps} and the
mask (2D). These window functions are assumed to be much sharper
than any variation in the power spectra. However such an assumption
is unlikely to pose a problem as the weak lensing power spectrum
lacks features unlike that of the CMB. The above expression is
expected to be reasonably accurate at high $l$ regime where the
noise dominates. The 3D harmonics that we have used in our
derivation based on the following definitions:

\beqa
&& [\sigma^{\sg,\sgp}(\oh)w(\oh)]_{lm}(k,k') \equiv \int [\sigma^{\sg,\sgp}(\oh,k,k')w(\oh)]Y^*_{lm}(\oh)d\oh  \nn\\
&& \sigma^{\sg,\sgp}(\oh,k,k') = {2 \over \pi} \int kdk j_l(kr) \int k'dk' j_l(k'r') \sigma^{\sg,\sgp}(\oh,r,r')
= {2 \over 2l +1 }  \sigma^{\sg,\sgp} \left [ \oh, {2l+1 \over k}, {2l+1 \over k'} \right ].
\label{noise_variance}
\eeqa

\n
In the final step we have used the Limber approximation. Similar terms such as $[\sigma^{\sg,\sgp}(\oh)w(\oh)]_{lm}(k,k')$
can be dealt with in a similar manner. In practice the evaluation of these terms will depend on the redshift
distribution of galaxies.

The deconvolution of the
error-covariance matrix can be performed using a similarity transformation.
It involves the mode-coupling matrix introduced before:

$\langle \delta \hat C_l^{\sg,\sgp}(k,k') \delta \hat C_{l'}^{\sg,\sgp}(k,k') \rangle =
M^{-1}_{lL} \langle\delta \tilde C_L^{\sg,\sgp}(k,k') \delta \tilde C_{L'}^{\sg,\sgp}(k,k') \rangle M^{-1}_{L'l'}$
A sum over repeated indices is assumed in this equation. The mode-coupling matrix
introduced here depends on the spins of the relevant fields involved $s$ and $s'$ Eq.(\ref{mode_coupling_cls}).
We will next extend this result to the skew spectrum
of arbitrary spinorial fields. For higher-order
spectra the results are more involved but they can be computed using the same
techniques considered here.

\subsection{Skew spectrum}

The expression for the skew spectrum, valid in the high-$l$ regime using the Limber approximation, is derived in Eq.(\ref{approx_bicls}).
The estimator for the skew spectrum quoted in Eq.(\ref{skew_cl}) depends on cross-correlating an arbitrary
product field $[_s\sg _{s'}\sg]_{lm}(k)$ against a third field $_{s''}\sgpp_{lm}(k')$.

In this section we compute the error-covariance
under certain simplifying assumptions. We will start from our definition of skew spectrum
Eq.(\ref{skew_cl}) and quote the result for the covariance, which can be derived using similar
procedure adopted for the derivation of covariance of the ordinary power spectrum.
The results correspond to three generic fields $\sg$, $\sgp$ and $\sgp$ with arbitrary spin weight
$s$, $s'$ and $s''=s+s'$ respectively:

\beqa
&& \langle \delta \myC_l^{\sg\sgp,\sgpp}(k,k')\; \delta \myC_{l'}^{\sg\sgp,\sgpp}(k,k')\rangle \approx
\Sigma^{SS}_{ll'}(k,k') + \Sigma^{SN}_{ll'}(k,k') + \Sigma^{NN}_{ll'}(k,k').
\eeqa

\n
The individual terms in terms of noise and signal power spectra are as follows:

\beqa
&& \Sigma^{SS}_{ll'}(k,k')= {1 \over 4\pi}\left \{ {\sqrt{\myC_l^{\sg\sgp,\sg\sgp}(k,k)\myC_{l'}^{\sg\sgp,\sg\sgp}(k',k')}
\sqrt{\myC_l^{\sgpp,\sgpp}(k,k)\myC_{l'}^{\sgpp,\sgpp}(k',k')} +
\myC_l^{\sg\sgp,\sgpp}(k,k)\myC_{l'}^{\sg\sgp,\sgpp}}(k',k') \right \} \nn \\
&& \quad\quad\quad\quad\quad\quad \times \sum_{l_am_a}
\left ( \begin{array} { c c c }
     l & l_a & l' \\
     s+s' & 0  & -(s+s')
  \end{array} \right )
\left ( \begin{array} { c c c }
     l & l_a & l' \\
     s'' & 0  & -s''
  \end{array} \right )
 |w_{l_a}|^2 \nn \\
&& \Sigma^{NN}_{ll'}(k,k') = {1 \over 4 \pi}\sum_{l_am_a}
\left ( \begin{array} { c c c }
     l & l_a & l' \\
     s+s' & 0 &  -(s+s')
  \end{array} \right )
\left ( \begin{array} { c c c }
     l & l_a & l' \\
     s'' & 0 &  -s''
  \end{array} \right )
\left \{ |w^2\sigma^{\sg\sg,\sg\sgp}|_{l_{\alpha}m_{\alpha}}(k)[w^2\sigma^{\sgpp,\sgpp}]^*_{l_{\alpha}m_{\alpha}}(k')
+ [w^2\sigma^{\sg\sgp,\sgpp}(k,k')]^2 \right \}
 \nn \\
&& \Sigma_{ll'}^{SN}(k,k') = {1 \over 4 \pi}\sum_{l_am_a}
\left ( \begin{array} { c c c }
     l & l_a & l' \\
     s+s' & 0 &  -(s+s')
  \end{array} \right )
\left ( \begin{array} { c c c }
     l & l_a & l' \\
     s'' & 0 &  -s''
  \end{array} \right )
\Big \{ 2~|w^2\sigma^{\sg\sgp,\sgpp}|_{l_{\alpha}m_{\alpha}}(k)|w^2|_{l_{\alpha}m_{\alpha}}\sqrt{\myC_l^{\sg\sgp,\sgpp}(k,k)\myC_{l'}^{\sg\sgp,\sgpp}(k,k)} \nn \\
&& \quad\quad + \; |w^2\sigma^{\sg\sgp,\sg\sgp}|_{l_{\alpha}m_{\alpha}}(k)|w^2|_{l_{\alpha}m_{\alpha}}\sqrt{\myC_l^{\sgpp,\sgpp}(k,k)\myC_{l'}^{\sgpp,\sgpp} (k,k)}
 + |w^2\sigma^{\sgpp,\sgpp}|_{l_{\alpha}m_{\alpha}}(k)|w^2|_{l_{\alpha}m_{\alpha}}\sqrt{\myC_l^{\sg\sgp,\sg\sgp}(k,k')\myC_{l'}^{\sg\sgp,\sg\sgp}(k,k')}
\Big \}
\eeqa

\n
The error-covariance depends on noise maps for the product field as well as the individual field.
The noise in our analysis is not assumed constant and can vary with position. The terms such as
$|w^2\sigma^{\sg\sg,\sg\sgp}|_{l_{\alpha}m_{\alpha}}(k)$ can also defined using expression similar to Eq.(\ref{noise_variance}).
It underlines the difficulty associated with an accurate error estimation beyond the power spectrum.

These results are based on various assumptions valid at high -$l$ regime. However for future, near
all-sky surveys for which the harmonic description is more appropriate these results can provide a good handle
on estimation errors. Alternatively the error-covariance can be computed using
Monte-carlo simulations. Simulating multiple copies of the observed sky, with all observational
details, can be computationally expensive. Hence, often simplifying assumptions
are employed to compute the covariance. The approach developed here can play a
complementary role in cross-checking and validating such results.
The lower-order covariance such as what
we have considered above typically depends on higher-order power spectra.
It is customary to quote error bars associated with estimated power spectra. However it is important to note that
the error-bars for the higher-order spectra such as the skew spectra
may not be fully informative, as the probability distribution of the skew spectrum
for a given $l$ can be skewed. In such cases, the error bars still can give
an idea of statistical scatter around the estimate.

The results for deconvolved PCL estimates for the skew spectrum can be computed using
a similarity transformation: $\langle \delta \hat C_l^{\sg\sgp,\sgpp}(k,k') \delta \hat C_{l'}^{\sg\sgp,\sgpp}(k,k') \rangle =
\sum_{LL'}G^{-1}_{lL} \langle\delta C_L^{\sg\sgp,\sgpp}(k,k') \delta C_{L'}^{\sg\sgp,\sgpp}(k,k') \rangle G^{-1}_{L'l'}$.
The mode-coupling matrix  $G$ is independent of the radial wave vector $k$ but depends on all three spin indices
Eq.(\ref{skew_mode_coupling_matrix}). Higher-order multi-spectra such as the the skew spectrum
are typically more correlated and binning may be needed for non-singular inversion.

It is also possible to compute the cross-covariance of these estimators for power spectrum and
the skew spectrum which can be used jointly. These can
result in tighter cosmological constraints. The results can be derived using the techniques presented above.
The coupling matrices are different for different spinorial fields. They depend on the spin
indices of the constituent spinorial fields. The spinorial fields considered above are however
completely generic. If we assume that the magentic or B-mode is absent then further simplification
can be achieved.

\subsection{Optimal Estimators}

The estimators that we have constructed can be generalised if we optimally weight the harmonics
with an inverse variance weight. The generalized two-to-one power spectrum $S_l^{\sg\sgp,\sgpp}$ in this case
takes the following form:
\beqa
&& \hat S_l^{\sg\sgp,\sgpp}(k,k') \equiv {1 \over 2l+1}\sum_{l'l''}{\Lambda}\;{ \hat B_{ll'l''}^{\sg\sgp\sgpp}(k,k,k')B_{ll'l''}^{\sg\sgp\sgpp}(k,k,k')}; \nn\\
&& {\Lambda^{-1}} = \left ( [C_l^{\sg\sg}(k,k)C_l^{\sgp\sgp}(k,k)C_l^{\sgpp\sgpp}(k',k')]^{-1}+ {\rm cyc.perm.} \right )
\eeqa

\n This particular result is valid for all-sky coverage. The
denominator $\Lambda$ is the scatter in the estimator in the
Gaussian limit. For partial sky coverage a more elaborate treatment
in line with \cite{MuHe09} is required. In the absence of spherical
symmetry due to lack of all-sky coverage or asymmetric noise, we
will have linear terms in the estimator. The estimator constructed
in this way can achieve maximum possible signal-to-noise for a given
data set. The one-point counterpart for this estimator, denoted as
$S$ can be recovered by summing over angular harmonics $l$, i.e.
$S^{\sg\sgp,\sgpp}(k,k')= \sum_l\;(2l+1)S_l^{\sg\sgp,\sgpp}(k,k')$.
An interesting point which we note here is that the hierarchical
ansatz is {\em factorizable} which will allow easy construction of
optimal weights. In general the expressions for gravity-induced the
bi- or trispectra are not factorisable. At bispectrum level we can
use the skew spectrum estimator to recover the tree amplitude $Q_3$
and similar estimators can be designed for the kurt spectra. The two
different kurt spectra will allow independent estimation of
topological amplitudes $R_a$ and $R_b$. However it is expected that
signal to noise at the level of trispectrum will be low. These
optimal estimators can be optimized to detect either the
gravity-induced non--Gaussianity or different models of primordial
non--Gaussianity. They can also be used to forecast
cross-contamination in a specific estimator from various sources of
non--Gaussianity.

It is worth repeating here that the next generation of weak lensing
surveys will have nearly  all-sky coverage. This will probe a wide
range of angular scales. The most commonly used technique for
statistical characterization of such surveys is real space analysis
using two- or three-point correlation functions. One of the
advantages of using the higher-order correlation hierarchy is its
ability to extract information from a complex survey geometry due to
partial sky coverage. However, the real space analysis introduces
highly correlated measurements for various angular scales. These
correlations are more dominant at small angular scales where most of
the observational information is contained. The alternative is to
use the harmonic space representation of the correlation functions,
e.g. the multi-spectra that are being studied here. Though
mathematically equivalent the power spectra or their higher-order
generalizations are much less used in the context of weak lensing.
However the theoretical interpretation of multispectra is much
simpler and different harmonics are much less correlated. The main
difficulty in harmonic analysis is related to the partial sky
coverage. Typically the mask consists of bright stars and saturated
spikes where no lensing measurements can be performed. The
analytical results presented here provide a general analysis of the
problem these pose. The results relate the convolved and deconvolved
power spectra that can be constructed from the higher-order
multispectra. The deconvolution process consists of simple matrix
inversion and can be performed for arbitrary sky coverage. For the
case of convergence the matrix representing mode-mode coupling in
the presence of mask is independent of the order of the
multi-spectra being probed. However that is not the case for the
case of shear or flexions. The formalism developed here also allows
for computation of scatter or variance associated with various
estimators.

\section{Conclusions}


It is now well accepted that the next generation of weak  lensing
surveys will play an important part in further reducing the
uncertainty in fundamental cosmological parameters, including those
that describe the evolution of equation of state of dark energy
\cite{Euclid}. They will also be instrumental in testing various
alternative gravity models (e.g.
\cite{HKV07,Amendola08,Benyon09,Schrabback09, Kilbinger09}). The
power of weak lensing surveys largely depends on the fact that they
can exploit both the angular diameter distance and the growth of
structure to constrain cosmological parameters. It is therefore very
important to develop analytical techniques and statistical tools
that can fully exploit the potential of future weak lensing surveys.


Typically without the redshift information the data from weak
lensing surveys are analyzed in projection for the entire source
distribution. However it was found that by binning sources in a few
photometric redshift bins the constraints improve \citep{Hu99}. In
recent years a full 3D formalism which exploits the photometric
redshifts of individual sources were developed. Such an approach
does not involve any binning; see \citet{Heav03,Castro05,HKT06}.
Further studies along these lines demonstrate that 3D lensing can
provide more powerful and tighter constraints on the dark energy
equation of state parameter, and on neutrino masses
\citep{deBernardis09, Jimenez10}, as well as testing braneworld and
other alternative gravity models. These constitute the main science
drivers for the future weak lensing surveys. Initial studies in weak
lensing focused on two-point correlation functions or the power
spectrum mainly due to the low signal-to-noise available for
higher-order studies from most first generation surveys. With the
availability of modern surveys it is useful to include the
non--Gaussianity information in the data analysis pipeline
\citep{TakadaJain04, Semboloni09} that can help to lift some of the
degeneracies in estimation of cosmological parameters involving
power spectrum alone.


In their recent work \cite{MuHeCo_wl1_10} have explored the possibility
of extending the higher-order statistics of convergence to 3D.
The main motivation of this work is to generalize those
results to spinorial objects and perform a
full 3D analysis for the higher-order statistics.
In this sense this is also an extension of results
derived in \cite{Mu_wl2_10} which analyzed higher-order statistics
of spinorial fields but only in projection (2D).
The results here are valid for all-sky surveys. It
depends on full 3D spherical harmonic decomposition
on the surface of the sky as well as along the radial
directions. Such an approach in analyzing  the data
from future surveys which will cover a large fraction
of the sky.


The higher-order statistics of convergence $\kappa$, shear $\gamma_{\pm}$ or
flexions $\cal F$ and $\cal G$ depend on accurate modelling of the
underlying density contrast $\delta$.  Various
models are used, such as the hierarchical ansatz which we use here, known to be valid
in the highly nonlinear regime. However the techniques developed
here are generic and can also be used in association
with other models such as the halo model.


The higher-order multispectra contain invaluable
information. Some of these information is however
degenerate because of symmetries associated with
higher-order correlation functions. It is difficult
to estimate the higher-order multispectra mode by mode
because of the associated scatter involved in such estimation
especially from a noisy data set. In \cite{MuHe09}, various
power spectra (skew spectrum, kurt spectra) were introduced, that are  associated
with a multispectra of a given order and can be
estimated in the presence of mask and noise.
These spectra carry some of the information
contents of the multispectra from which they are constructed. In our present
study we express the skew spectrum and two degenerate kurt spectra
of generic spinorial fields in terms of the bi- and trispectrum.
This extends earlier results for the convergence (spin-$0$)
field. Extending the previously introduced
3D power spectrum $C_l(k_1,k_2)$ to higher-order, we introduce
a series of power spectra related to multispectra at each order.
We have introduced the {\em 3D skew spectrum} $C_l^{\sg\sgp,\sgpp}(k_1,k_2)$
associated with the bispectrum of arbitrary triplets of spinorial fields $\sg,\sg,\sgpp$.
Analogously, at the level of trispectrum we have introduced two 3D kurt spectra
$C_l^{\sg\sgp\sgpp,\sgppp}(k_1,k_2)$ and $C_l^{\sg\sgp,\sgpp\sgppp}(k_1,k_2)$  for arbitrary choice of
spinorial fields. These extends the skew- and kurt spectra defined
in \cite{MuHeCo_wl1_10} where harmonic decomposition was performed only on the
surface of the celestial sphere and a real space analysis was performed on
the radial direction leading to a mixed representation of skew spectrum
$C_l^{(2,1)}(r_2,r_1)$ as well as their higher-order counterparts i.e. the two kurt spectra
$C_l^{(2,2)}(r_2,r_1)$ and $C_l^{(3,1)}(r_2,r_1)$.


The generic expression for the skew- and kurt spectra involve
spherical Bessel functions. We simplified these radial integrals by
using the Limber approximation, whose accuracy scales as ${\cal
O}({1 \over l^4})$. We show that at each order the Limber
approximation can reduce the dimensionality of the integrals to
unity which dramatically reduces computational cost. Both the Limber
approximation and the hierarchical ansatz are accurate at smaller
scales and their joint use can help us to compute the skew- and
kurt- spectra very efficiently with reasonable accuracy, but the
method can accommodate different models for nonlinear clustering.


We also present analytical results for dealing with a mask, via a
pseudo-$C_l$ approach, encapsulated in a mode-mixing matrix. The
estimation of unbiased skew- or kurt spectra are done by simple
inversion of the mixing matrix $M$, which depends on the spins
associated with the spinorial fields.   Some regularisation will
normally be required.  The presence of an observational mask
typically only induces mode-mixing on the celestial sphere and not
on the radial direction. We have also showed how our formalism
presented here can be used also for the computation of scatter under
certain simplifying assumptions in the presence of an observational
mask, and we have identified individual terms that correspond to
contributions from noise, partial sky coverage (cosmic variance) and
cross terms.


The results presented here will be relevant for the study of cosmic
magnification studies in 3D as well as in many other contexts where
integrated radial information is used. The estimators for skew or
kurt spectra that we have described here can be improved by inverse
variance weighting of 3D harmonics. Finally, to summarize:

\begin{itemize}
\item{We have studied higher-order multispectra in the context of 3D weak lensing surveys.}
\item{We use a full 3D Fourier decomposition which employ spin-weight spherical harmonics.}
\item{Our generic results are valid for arbitrary 3D spinorial objects.}
\item{The results are relevant for convergence $\kappa$,
magnification $\mu$, shear $\gamma_{\pm}$ as well as flexions $\cal F$ and $\cal G$ or an arbitrary
scalar tracer field $\Phi$.}
\item{In our analysis we define power spectra $C_l(k_1,k_2)$ that are related to the bispectrum (skew spectra)
  and to the trispectrum (kurt spectra).}
\item{We provide both all-sky exact results and corresponding approximate results using the Limber approximation.}
\item{Use of Limber's approximation reduces multidimensional integrations along the radial direction to
 one-dimensional integrals.}
\item{We show how the multi-spectra can be recovered from a masked sky in the presence of noise, and show how the presence of masks mixes modes not only on the surface of the sky but in the radial direction.}
\item{The modelling was done using the hierarchical ansatz but the formalism can work with any input
underlying density multispectra.}
\item{Under certain simplifying approximations, we also obtain expressions for the covariance of our
power spectra and skew spectra estimators.}
\item{We outline how inverse variance weights can be introduced and optimal estimators can be defined
for the detection of a specific type of non--Gaussianity.}
\item{The formalism can be relevant in many other contexts where line-of-sight integrations
of non-Gaussianities are performed or in studies involving cross-spectra or mixed-bispectra.}
\end{itemize}

In this paper we have ignored many observational complexities for simplicity, such as
that in a realistic survey the lensing potential can only be sampled
at the discrete positions of galaxies, and the average number of source galaxies will decline with redshift. We also
ignore photometric redshift errors.
\section{Acknowledgements}
\label{acknow} DM and PC acknowledge support from STFC standard
grant ST/G002231/1 at the School of Physics and Astronomy at Cardiff
University, where this work was completed. It is a pleasure for DM to
acknowledge useful discussions with Ludovic van Waerbeke and Joseph Smidt.
TK is supported by STFC rolling grant number RA0888.

\bibliography{3dshearflexion_PC.bbl}

\appendix

\section{Useful Mathematical Relations}

\subsection{Spherical Bessel Functions}

\n
The orthogonality relationship for the spherical Bessel functions is given by the following expression:

\beqa
\int k^2 j_l(kr_1) j_l(kr_2) dk = \left [ {\pi \over 2 r_1^2} \right ] \delta_{1D}(r_1-r_2).
\label {eq:limber_approx1}
\eeqa

\n
The extended Limber approximation is also implemented through the following approximate relation \cite{LoAf08}:

\beqa
\int k^2 F(k) j_l(kr_1) j_l(kr_2) dk \sim  \left [ {\pi \over 2 r_1^2} \right ] F \left ( {l \over r_1} \right ) \delta_{1D}(r_1-r_2).
\label{eq:limber_approx2}
\eeqa

\n Thus, for high $l$, the spherical Bessel functions can be
replaced by a Dirac delta function $\delta_{1D}$:

\beqa
\displaystyle \lim_{x\to\infty} j_l(x) = \sqrt {\pi \over 2l+1} \delta_{1D} \left ( l+ {1 \over 2} - x\right ).
\label{eq:limber_approx3}
\eeqa

\subsection{3j Symbols}

\n
The following properties of $3j$ symbols were used to simplify various expressions.

\beqa
\sum_{l_3m_3} (2l_3+1) \left ( \begin{array}{ c c c }
     l_1 & l_2 & l_3 \\
     m_1 & m_2 & m_3
  \end{array} \right )
\left ( \begin{array}{ c c c }
     l_1 & l_2 & l_3 \\
     m_1' & m_2' & m_3
  \end{array} \right ) = \delta^K_{m_1m_1'} \delta^K_{m_2m_2'}
\eeqa

\beqa
\sum_{m_1m_2} \left ( \begin{array}{ c c c }
     l_1 & l_2 & l_3 \\
     m_1 & m_2 & m_3
  \end{array} \right)
\left ( \begin{array}{ c c c }
     l_1 & l_2 & l_3' \\
     m_1 & m_2 & m_3'
  \end{array} \right) = {\mathcal \delta^K_{l_3l_3'} \delta^K_{m_3m_3'} \over 2l_3 + 1}
\eeqa

\end{document}